\documentclass[a4paper,11pt]{article}
\usepackage{amssymb}

\usepackage{amsmath}
\usepackage[utf8]{inputenc}
\usepackage{fullpage}
\usepackage{boxedminipage}
\usepackage{listings}
\usepackage{minitoc}
\usepackage[pdftex]{graphicx}
\usepackage{graphicx}

\makeatletter \@addtoreset{equation}{section} \makeatother
\renewcommand{\theequation}{\thesection.\arabic{equation}}
\newif\ifpdf \ifx\pdfoutput\undefined \pdffalse
\else \pdfoutput=1 \pdftrue \fi \ifpdf \else \fi
\input epsf

\begin{document}

\ifpdf\DeclareGraphicsExtensions{.pdf, .jpg, .tif} \else%
\DeclareGraphicsExtensions{.eps, .jpg} \fi
\begin{titlepage}

    \thispagestyle{empty}
    \begin{flushright}
         \hfill{CERN-PH-TH/2009-070} \\
        \hfill{UCLA/09/TEP/51}\\
        \hfill{SU-ITP-09/19}\\
    \end{flushright}

    \vspace{2pt}
    \begin{center}
        { \Huge{\bf SAM Lectures\\on Extremal Black Holes\\
        \vspace{10pt}in $d=4$ Extended Supergravity}}

        \vspace{8pt}

        {\large{\bf Stefano Bellucci$^\clubsuit$, Sergio Ferrara$^{\diamondsuit\clubsuit\flat}$,\\Murat G\"{u}naydin$^{\spadesuit}$\vspace{5pt} and \ Alessio Marrani$^{\heartsuit}$}}

        \vspace{5pt}

        {$\clubsuit$ \it INFN - Laboratori Nazionali di Frascati, \\
        Via Enrico Fermi 40,00044 Frascati, Italy\\
        \texttt{bellucci@lnf.infn.it}}

        \vspace{2pt}

        {$\diamondsuit$ \it Physics Department,Theory Unit, CERN, \\
        CH 1211, Geneva 23, Switzerland\\
        \texttt{sergio.ferrara@cern.ch}}

        \vspace{2pt}

        {$\flat$ \it Department of Physics and Astronomy,\\
        University of California, Los Angeles, CA USA}

        \vspace{2pt}

        {$\spadesuit$ \it Department of Physics, Penn State University\\
        University Park, PA16802, USA\\
        \texttt{murat@phys.psu.edu}}

        \vspace{1pt}

        {$\heartsuit$ \it Stanford Institute for Theoretical Physics\\
        Department of Physics, 382 Via Pueblo Mall, Varian Lab,\\
        Stanford University, Stanford, CA 94305-4060, USA\\
        \texttt{marrani@lnf.infn.it}}

        \vspace{15pt}
        \noindent \textit{Contribution to the Proceedings of the\\School on Attractor Mechanism 2007 (SAM2007),\\June 18--22 2007, INFN--LNF, Frascati, Italy}
\end{center}

\vspace{5pt}

\begin{abstract}
We report on recent results in the study of extremal black hole
attractors in $N=2$, $d=4$ ungauged Maxwell-Einstein supergravities.

For homogeneous symmetric scalar manifolds, the three general
classes of attractor solutions with non-vanishing Bekenstein-Hawking
entropy are discussed. They correspond to three
(inequivalent) classes of orbits of the charge vector, which sits in the relevant symplectic representation $R_{V}$ of the $U$%
-duality group. Other than the $\frac{1%
}{2}$-BPS one, there are two other distinct non-BPS classes of
charge orbits, one of which has vanishing central charge.

The complete classification of the $U$%
-duality orbits, as well as of the moduli spaces of non-BPS
attractors (spanned by the scalars which are not stabilized at the
black hole event horizon), is also reviewed.

Finally, we consider the analogous classification for $N\geqslant
3$-extended, $d=4$ ungauged supergravities, in which also the
$\frac{1}{N}$-BPS attractors yield a related moduli space.

\end{abstract}

\end{titlepage}
\newpage\tableofcontents
\def\theequation{\arabic{section}.\arabic{equation}}
\section{\label{Intro}Introduction}

In the framework of \textit{ungauged} Einstein supergravity theories in $d=4$
space-time dimensions, the fluxes of the two-form electric-magnetic field
strengths determine the charge configurations of stationary, spherically
symmetric, asymptotically flat extremal black holes (BHs). Such fluxes sit
in a representation $R_{V}$ of the $U$-duality\footnote{%
Here $U$-duality is referred to as the ``continuous'' version, valid
for large values of the charges, of the $U$-duality groups
introduced by Hull and Townsend \cite{HT}.} group $G_{4}$ of the
underlying $d=4$ supergravity, defining the embedding of $G$ into
the larger symplectic group  $Sp\left( 2n, \mathbb{R}\right) $.
Moreover, after the study of \cite{FG1}, for \textit{symmetric}
scalar manifolds $\frac{G_{4}}{H_{4}}$ (see Eq. (\ref{M4}) below)
the fluxes
belong to distinct classes of orbits of the representation $R_{V}$, \textit{%
i.e.} the $R_{V}$-representation space of $G_{4}$ is actually \textit{%
``stratified''} into disjoint classes of orbits. Such orbits are defined and
classified by suitable constraints on the (lowest order, actually unique) $G$%
-invariant $\mathcal{I}$ built out of the symplectic representation $R_{V}$.

For all $N\geqslant 3$, $d=4$ supergravities the scalar manifold of the
theory is an homogeneous \textit{symmetric} space $\frac{G_{4}}{H_{4}}$.
Thus, for such theories some relations between the coset expressions of the
aforementioned orbits and different real (non-compact) forms of the
stabilizer $H_{4}$ can be established \cite{BFGM1}. It is here worth
remarking that the \textit{``large''} charge orbits (having $\mathcal{I}\neq
0$) support the \textit{Attractor Mechanism }\cite{FKS}-\nocite
{Strom,FK1,FK2}\cite{FGK}, whereas the \textit{``small''} ones (having $%
\mathcal{I}=0$) do not.

Recently, a number of papers have been devoted to the investigation of
\textit{extremal }BH \textit{attractors} (see \textit{e.g.} \cite{Sen-old1}%
-- \nocite{GIJT,Sen-old2,K1,TT,G,GJMT,Quantum-1,Ebra1,K2,Ira1,Tom,
BFM,AoB-book,FKlast,FG2,Ebra2,rotating-attr,K2-bis,Misra1,Lust2,Morales,
BFMY,Astefa,CdWMa,
DFT07-1,BFM-SIGRAV06,Cer-Dal,ADFT-2,Saraikin-Vafa-1,
Ferrara-Marrani-1,TT2,
ADOT-1,ferrara4,Astefanesei,CCDOP,Misra2,Quantum-2,Ceresole,
Anber,Myung1,BMOS-1,Hotta,
Gao,PASCOS07,Sen-review,Belhaj1,AFMT1,Gaiotto1,
BFMS1,GLS1,ANYY1,review-Kallosh,Cai-Pang,Vaula,
Li,BFMY2,Saidi2,Saidi3,Saidi4,Sen-Entropy-1,
Attractors-Black,FHM,Unattractor,Vaula2,Trigiante,Wall-Crossing,
Gnecchi-1,
stu-unveiled,Chiral-Ring,Sen-Entropy-2,Quantum-Lift,FMMS-1,Hotta-2,CFT-Duals,BNS-1,Sen-Entropy-3,Trigiante-2,
Fre-Sorin-1,Fre-Sorin-2,Fre-Sorin-Chemissany-1} \cite{CFGM-1}; for
further developments and Refs., see also \textit{e.g.}
\cite{OSV}--\nocite{OVV,ANV,GSV}\cite{Pioline-review}), essentially
because new classes of solutions to the so-called \textit{Attractor
Equations} were
(re)discovered. Such new solutions have been found to determine non-BPS (%
\textit{Bogomol'ny-Prasad-Sommerfeld}) BH horizon geometries, breaking
\textit{all} supersymmetries (\textit{if any}).\medskip

The present report, originated from lectures given at the
\textit{School on Attractor Mechanism} (\textit{SAM2007}), held on
June 18-22 2007 at INFN National Laboratories in Frascati (LNF),
Italy, is devoted to an introduction to the foundations of the
theory of $U$-duality orbits in the theory of extremal BH attractors
in $N=2$, $d=4$ MESGT's based on symmetric manifolds. Also
$N\geqslant 3$-extended, $d=4$ supergravities, as well as the issue
of moduli spaces of attractor solutions, will be briefly considered.
Our review incorporates some of the more recent developments that
have taken place since \textit{SAM2007}.

\bigskip

The plan of the report is as follows.\smallskip

Sect. \ref{N=2,d=4,Symm} is devoted to the treatment of $N=2$, $d=4$ MESGT's
based on symmetric scalar manifolds.

In Subsect. \ref{N=2-Orbits} we review some basic facts about such theories,
concerning their \textit{``large''} charge orbits and the relations with the
extremal BH solutions to the corresponding Attractor Eqs..

Thence, Subsect. \ref{N=2-Attractors} reports the general analysis,
performed in \cite{BFGM1}, of the three classes of extremal BH
attractors of $N=2$, $d=4$ symmetric \textit{magic} \footnote{These
theories were called \textit{"magical"} MESGT's in the original
papers. In some of the recent literature they are referred to as
\textit{"magic"} MESGT's which we shall adopt in this review.}
MESGT's, and of the corresponding classes of \textit{``large''}
charge orbits in the symplectic representation
space of the relevant $d=4$ $U$-duality group. In particular, the $\frac{1}{2%
}$-BPS solutions are treated in Subsubsect. \ref{N=2-Attractors-BPS}, while
the two general species of non-BPS $Z\neq 0$ and non-BPS $Z=0$ attractors
are considered in Subsubsect. \ref{N=2-Attractors-non-BPS}.

The splittings of the mass spectra of $N=2$, $d=4$ symmetric \textit{magic}
MESGT's along their three classes of \textit{``large''} charge orbits \cite
{BFGM1}, and the related issues of massless Hessian modes and moduli spaces
of attractor solutions, are considered\footnote{%
In the present report we do not consider the other $N=2$, $d=4$ MESGT's with
symmetric scalar manifolds, given by the two infinite sequences $\frac{%
SU(1,1+n)}{U(1)\times SU(1+n)}$ and $\frac{SU(1,1)}{U(1)}\times \frac{%
SO(2,2+n)}{SO(2)\times SO(2+n)}$. These theories are treated in detail in
the two Appendices of \cite{BFGM1}.} in Subsect. \ref{N=2-Spectra}.

Subsect. \ref{Moduli-Spaces} deals with the crucial result that the massless
Hessian modes of the effective BH potantial $V_{BH}$ of $N=2$, $d=4$ MESGT's
based on symmetric scalar manifolds at its critical points actually
correspond to \textit{``flat''} directions. Such \textit{``flat''}
directions are nothing but the scalar degrees which are not stabilized at
the event horizon of the considered $d=4$ extremal BH, thus spanning a
\textit{moduli space} associated to the considered attractor solution.
Nevertheless, the BH entropy is still well defined, because, due to the
existence of such \textit{``flat''} directions, it is actually \textit{%
independent} on the unstabilized scalar degrees of freedom.

Actually, moduli spaces of attractors solutions exist \textit{at least} for
all \textit{ungauged} supergravities based on homogeneous scalar manifolds.
The classification of such moduli spaces (and of the corresponding
supporting \textit{``large''} orbits of $U$-duality for $N\geqslant 3$, $d=4$
supergravities is reported in Sect. \ref{N>2,d=4}.

Sect. \ref{Conclusions} concludes the present report, with some final
comments and remarks.

\section{\label{N=2,d=4,Symm}$N=2$, $d=4$ Symmetric MESGT's}
\def\theequation{2.\arabic{subsection}.\arabic{equation}}
\subsection{\label{N=2-Orbits}$U$-Duality \textit{``Large''} Orbits}

The critical points of the BH effective potential $V_{BH}$ for all $N=2$
symmetric special geometries in $d=4$ are generally referred to as
attractors. These extrema describe the \textit{``large''} configurations
(BPS as well as non-BPS) of $N=2,6,8$ supergravities, corresponding to a
finite, non-vanishing quartic invariant $\mathcal{I}_{4}$ and thus to
extremal BHs with classical non-vanishing entropy $S_{BH}\neq 0$ . The
related orbits in the $R_{V}$ of the $d=4$ $U$-duality group $G_{4}$ will
correspondingly be referred to as \textit{``large''} orbits. The attractor
equations for BPS configurations were first studied in \cite{FKS}-\nocite
{Strom,FK1}\cite{FK2}, and flow Eqs. for the general case were given in \cite
{FGK}.

Attractor solutions and their \textit{``large''} charge orbits in $d=5$ have
been recently classified for the case of all rank-2 symmetric spaces in \cite
{FG2}.\smallskip

In \cite{BFGM1} the results holding for $N=8$, $d=4$ supergravity were
obtained also for the particular class of $N=2$, $d=4$ symmetric
Maxwell-Einstein supergravity theories (MESGT's) \cite{GST1,GST2,GST3},
which we will now review. Such a class consists of $N=2$, $d=4$
supergravities sharing the following properties:

$i)$ beside the supergravity multiplet, the matter content is given only by
a certain number $n_{V}$ of Abelian vector multiplets;

$ii)$ the space of the vector multiplets' scalars is an homogeneous
symmetric special K\"{a}hler manifold, i.e. a special K\"{a}hler manifold
with coset structure
\begin{equation}
\frac{G_{4}}{H_{4}}\equiv \frac{G}{H_{0}\times U(1)},  \label{M4}
\end{equation}
where $G\equiv G_{4}$\ is a semisimple non-compact Lie group and $%
H_{4}\equiv H_{0}\times U(1)$\ is its maximal compact subgroup (\textit{mcs}%
) (with symmetric embedding, as understood throughout);

$iii)$ the charge vector in a generic (dyonic) configuration with $n_{V}+1$
electric and $n_{V}+1$ magnetic charges sits in a real (symplectic)
representation $R_{V}$ of $G$ of $dim\left( R_{V}\right) =2\left(
n_{V}+1\right) $.

By exploiting such special features and relying on group theoretical
considerations, in \cite{BFGM1} the coset expressions of the various
distinct classes of \textit{``large''} orbits (of dimension $2n_{V}+1$) in
the $R_{V}$-representation space of $G$ were related to different real
(non-compact) forms of the compact group $H_{0}$. Correspondingly, the $N=2$%
, $d=4$ Attractor Eqs. were solved for all such classes, also studying the
scalar mass spectrum of the theory corresponding to the obtained
solutions.\medskip

The symmetric special K\"{a}hler manifolds of $N=2$, $d=4$ MESGT's have been
classified in the literature (see \textit{e.g.} \cite{CVP,dWVVP} and Refs.
therein). \textbf{\ }All such theories can be obtained by dimensional
reduction of the $N=2$, $d=5$ MESGT's that were constructed in \cite
{GST1,GST2,GST3}. The MESGT's with symmetric manifolds that originate from $%
d=5$ all have cubic prepotentials determined by the norm form of the Jordan
algebra of degree three that defines them \cite{GST1,GST2,GST3}.

The unique exception is provided by the infinite sequence ($n\in \mathbb{N}%
\cup \left\{ 0\right\} $, $n_{V}=n+r=n+1$) \cite{Luciani}
\begin{equation}
I_{n}:\text{\ }\frac{SU(1,1+n)}{U(1)\times SU(1+n)},\text{ }r=1,
\label{quadr}
\end{equation}
where $r$ stands for the \textit{rank} of the coset throughout. This is
usually referred to as \textit{minimal coupling} sequence, and it is endowed
with quadratic prepotential. It should be remarked that the $N=2$ \textit{%
minimally coupled} supergravity is the only (symmetric) $N=2$, $d=4$ MESGT
which yields the \textit{pure }$N=2$ supergravity simply by setting $n=-1$
(see \textit{e.g.} \cite{Gnecchi-1} and Refs. therein).

Only another infinite symmetric sequence exists, namely ($n\in \mathbb{N}%
\cup \left\{ 0,-1\right\} $, $n_{V}=n+r=n+3$)
\begin{equation}
II_{n}:\frac{SU(1,1)}{U(1)}\times \frac{SO(2,2+n)}{SO(2)\times SO(2+n)},r=3,
\label{cub}
\end{equation}
This one has a $d=5$ origin and its associated Jordan algebras are not
simple. It is referred to as the \textit{``generic Jordan family''} since it
exists $\forall n\in \mathbb{N}\cup \left\{ 0,-1\right\} $. The first
elements of such sequences (\ref{quadr}) and (\ref{cub}) correspond to the
following manifolds and holomorphic prepotential functions in special
coordinates:
\begin{gather}
I_{0}:\frac{SU(1,1)}{U(1)},\text{ }F(t)=-\frac{i}{2}\left( 1-t^{2}\right) ;
\\
\notag \\
II_{-1}:\frac{SU(1,1)\times SO(2,1)}{U(1)\times SO(2)}=\left( \frac{SU(1,1)}{%
U(1)}\right) ^{2},F(s,t)=st^{2}; \\
\notag \\
II_{0}:\frac{SU(1,1)\times SO(2,2)}{U(1)\times SO(2)\times SO(2)}=\left(
\frac{SU(1,1)}{U(1)}\right) ^{3},F(s,t,u)=stu;  \label{stu}
\end{gather}
It is here worth remarking that the so-called $t^{3}$ model, corresponding
to the following manifold and holomorphic prepotential function in special
coordinates:
\begin{equation}
\frac{SU(1,1)}{U(1)},\text{ }F(t)=t^{3},
\end{equation}
is an \textit{isolated case} in the classification of symmetric SK manifolds
(see \textit{e.g.} \cite{CFG}; see also \cite{LA08-Proc} and Refs. therein),
but it can be thought also as the ``$\mathit{t}^{3}$ \textit{degeneration'' }%
of the $stu$ model (see \textit{e.g.} \cite{BMOS-1}; see also Subsect. \ref
{Moduli-Spaces} for a treatment of models $I_{0}$ and $t^{3}$).

As mentioned, all manifolds of type $I$ correspond to quadratic
prepotentials ($C_{ijk}=0$), and all manifolds of type $II$ correspond to
cubic prepotentials (in special coordinates $F=\frac{1}{3!}%
d_{ijk}t^{i}t^{j}t^{k}$ and therefore $C_{ijk}=e^{K}d_{ijk}$, where $K$
denotes the K\"{a}hler potential and $d_{ijk}$ is a completely symmetric
rank-3 constant tensor). The $3$-moduli case $II_{0}$ is the well-known $stu$
model \cite{DLR,BKRSW} (see also \textit{e.g.} \cite{stu-unveiled}\ and
Refs. therein), whose noteworthy \textit{triality symmetry} has been
recently related to \textit{quantum information theory} \cite
{Duff-Cayley,KL,Levay-QIT,Ferrara-Duff-QIT,Duff-group-QIT,Levay-et-al-QIT}.

\textbf{\ }Beside the infinite sequence $II$, there exist four other MESGT's
defined by simple Euclidean Jordan algebras of degree three with the
following rank-$3$ symmetric manifolds:
\begin{eqnarray}
&&
\begin{array}{l}
III:\frac{E_{7(-25)}}{E_{6}\times U(1)};
\end{array}
\\
&&  \notag \\
&&
\begin{array}{l}
IV:\frac{SO^{\ast }(12)}{U(6)};
\end{array}
\\
&&  \notag \\
&&
\begin{array}{l}
V:\frac{SU(3,3)}{S\left( U(3)\times U(3)\right) }=\frac{SU(3,3)}{SU(3)\times
SU(3)\times U(1)};
\end{array}
\\
&&  \notag \\
&&
\begin{array}{l}
VI:\frac{Sp(6,\mathbb{R})}{U(3)}.
\end{array}
\end{eqnarray}
The $N=2$, $d=4$ MESGT's whose geometry of scalar fields is given by the
manifolds $III$-$VI$ are called \textit{``magic''}, since their symmetry
groups are the groups of the famous Magic Square of Freudenthal, Rozenfeld
and Tits associated with some remarkable geometries \cite{Freudenthal2,magic}%
. The four $N=2$, $d=4$ \textit{magic} MESGT's $III$-$VI$, as their $d=5$
versions, are defined by four simple Euclidean Jordan algebras $J_{3}^{%
\mathbb{O}}$, $J_{3}^{\mathbb{H}}$, $J_{3}^{\mathbb{C}}$ and $J_{3}^{\mathbb{%
R}}$ of degree $3$ with irreducible norm forms, namely by the Jordan
algebras of Hermitian $3\times 3$ matrices over the four division algebras,
\textit{i.e.} respectively over the octonions $\mathbb{O}$, quaternions $%
\mathbb{H}$, complex numbers $\mathbb{C}$ and real numbers $\mathbb{R}$ \cite
{GST1,GST2,GST3,Jordan,Jacobson,Guna1,GPR}.

By denoting with $n_{V}$ the number of vector multiplets coupled to the
supergravity one, the total number of Abelian vector fields in the
considered $N=2$, $d=4$ MESGT is $n_{V}+1$; correspondingly, the real
dimension of the corresponding scalar manifold is $2n_{V}=dim\left( G\right)
-dim\left( H_{0}\right) -1$. Since the $2\left( n_{V}+1\right) $-dim. vector
of extremal BH charge configuration is given by the fluxes of the electric
and magnetic field-strength two-forms, it is clear that $dim_{\mathbb{R}%
}\left( R_{V}\right) =2\left( n_{V}+1\right) $.

Since $H_{0}$ is a proper compact subgroup of the duality semisimple group $%
G $, one can decompose the $2\left( n_{V}+1\right) $-dim. real symplectic
representation $R_{V}$ of $G$ in terms of complex representations of $H_{0}$%
, obtaining in general the following decomposition scheme:
\begin{equation}
R_{V}\longrightarrow R_{H_{0}}+\overline{R_{H_{0}}}+\mathbf{1}_{\mathbb{C}}+%
\overline{\mathbf{1}}_{\mathbb{C}}=R_{H_{0}}+\mathbf{1}_{\mathbb{C}}+c.c.,
\label{decomp1}
\end{equation}
where ``$c.c.$'' stands for the complex conjugation of representations
throughout, and $R_{H_{0}}$ is a certain complex representation of $H_{0}$.%
\textbf{\ }

The basic data of the cases $I$-$VI$ listed above are summarized in Tables 1
and 2.
\begin{table}[t]
\begin{center}
\begin{tabular}{|c||c|c|}
\hline
& $I$ & $II$ \\ \hline\hline
$G$ & $SU(1,1+n)$ & $SU(1,1)\times SO(2,2+n)$ \\ \hline
$H_{0}$ & $SU(1+n)$ & $SO(2)\times SO(2+n)$ \\ \hline
$r$ & $1$ & $3$ \\ \hline
$dim_{\mathbb{R}}\left( \frac{G}{H_{0}\times U(1)}\right) $ & $2\left(
n+1\right) $ & $2\left( n+3\right) $ \\ \hline
$n_{V}$ & $n+r=n+1$ & $n+r=n+3$ \\ \hline
$R_{V}$ & $\left( \mathbf{2}\left( \mathbf{n+2}\right) \right) _{\mathbb{R}}$
& $\left( \mathbf{2}\left( \mathbf{n+4}\right) \right) _{\mathbb{R}}$ \\
\hline
$R_{H_{0}}$ & $\left( \mathbf{n+1}\right) _{\mathbb{C}}$ & $\left( \mathbf{%
n+2+1}\right) _{\mathbb{C}}$ \\ \hline
$dim_{\mathbb{R}}\left( R_{V}\right) $ & $2\left( n+2\right) $ & $2\left(
n+4\right) $ \\ \hline
$dim_{\mathbb{R}}\left( R_{H_{0}}\right) $ & $2\left( n+1\right) $ & $%
2\left( n+3\right) $ \\ \hline
$
\begin{array}{c}
R_{V} \\
\downarrow \\
R_{H_{0}}+\mathbf{1}_{\mathbb{C}}+ \\
+c.c.
\end{array}
$ & $
\begin{array}{c}
\left( \mathbf{2}\left( \mathbf{n+2}\right) \right) _{\mathbb{R}} \\
\downarrow \\
\left( \mathbf{n+1}\right) _{\mathbb{C}}+\mathbf{1}_{\mathbb{C}}+ \\
+c.c.
\end{array}
$ & $
\begin{array}{c}
\left( \mathbf{2}\left( \mathbf{n+4}\right) \right) _{\mathbb{R}} \\
\downarrow \\
\left( \mathbf{n+2+1}\right) _{\mathbb{C}}+\mathbf{1}_{\mathbb{C}}+ \\
+c.c.
\end{array}
$ \\ \hline
\end{tabular}
\end{center}
\caption{\textbf{Data of the two sequences of symmetric $N=2$, $d=4$ MESGT's}
}
\end{table}
\begin{table}[h]
\begin{center}
\begin{tabular}{|c||c|c|c|c|}
\hline
& $III:J_{3}^{\mathbb{O}}$ & $IV:J_{3}^{\mathbb{H}}$ & $V:J_{3}^{\mathbb{C}}$
& $VI:J_{3}^{\mathbb{R}}$ \\ \hline\hline
$G$ & $E_{7(-25)}$ & $SO^{\ast }(12)$ & $SU(3,3)$ & $Sp(6,\mathbb{R})$ \\
\hline
$H_{0}$ & $E_{6}$ & $SU(6)$ & $SU(3)\times SU(3)$ & $SU(3)$ \\ \hline
$r$ & $3$ & $3$ & $3$ & $3$ \\ \hline
$dim_{\mathbb{R}}\left( \frac{G}{H_{0}\times U(1)}\right) $ & $54$ & $30$ & $%
18$ & $12$ \\ \hline
$n_{V}$ & $27$ & $15$ & $9$ & $6$ \\ \hline
$R_{V}$ & $\mathbf{56}_{\mathbb{R}}$ & $\mathbf{32}_{\mathbb{R}}$ & $\mathbf{%
10}_{\mathbb{R}}$ & $\mathbf{14}_{\mathbb{R}}^{\prime }$ \\ \hline
$R_{H_{0}}$ & $\mathbf{27}_{\mathbb{C}}$ & $\mathbf{15}_{\mathbb{C}}$ & $%
\left( \mathbf{3,3}^{\prime }\right) _{\mathbb{C}}$ & $\mathbf{6}_{\mathbb{C}%
}$ \\ \hline
$dim_{\mathbb{R}}\left( R_{V}\right) $ & $56$ & $32$ & $20$ & $14$ \\ \hline
$dim_{\mathbb{R}}\left( R_{H_{0}}\right) $ & $54$ & $30$ & $18$ & $12$ \\
\hline
$
\begin{array}{c}
R_{V} \\
\downarrow \\
R_{H_{0}}+\mathbf{1}_{\mathbb{C}}+ \\
+c.c.
\end{array}
$ & $
\begin{array}{c}
\mathbf{56}_{\mathbb{R}} \\
\downarrow \\
\mathbf{27}_{\mathbb{C}}+\mathbf{1}_{\mathbb{C}}+ \\
+c.c.
\end{array}
$ & $
\begin{array}{c}
\mathbf{32}_{\mathbb{R}} \\
\downarrow \\
\mathbf{15}_{\mathbb{C}}+\mathbf{1}_{\mathbb{C}}+ \\
+c.c.
\end{array}
$ & $
\begin{array}{c}
\mathbf{10}_{\mathbb{R}} \\
\downarrow \\
\left( \mathbf{3,3}^{\prime }\right) _{\mathbb{C}}+\mathbf{1}_{\mathbb{C}}+
\\
+c.c.
\end{array}
$ & $
\begin{array}{c}
\mathbf{14}_{\mathbb{R}}^{\prime } \\
\downarrow \\
\mathbf{6}_{\mathbb{C}}+\mathbf{1}_{\mathbb{C}}+ \\
+c.c.
\end{array}
$ \\ \hline
\end{tabular}
\end{center}
\caption{\textbf{Data of the four \textit{magic} symmetric $N=2$, $d=4$
MESGT's}. {$\mathbf{14}_{\mathbb{R}}^{\prime }$ is the\textbf{\ }rank-}${3}${%
\ antisymmetric tensor representation of $Sp(6,\mathbb{R})$.} In $\left(
\mathbf{3,3}^{\prime }\right) _{\mathbb{C}}$ the prime distinguishes the
representations of the two distinct $SU(3)$ groups}
\end{table}

It was shown in \cite{FG1} that $\frac{1}{2}$-BPS orbits of $N=2$, $d=4$
symmetric MESGT's are coset spaces of the form
\begin{equation}
\begin{array}{l}
\mathcal{O}_{\frac{1}{2}-BPS}=\frac{G}{H_{0}}, \\
\\
dim_{\mathbb{R}}\left( \mathcal{O}_{\frac{1}{2}-BPS}\right) =dim\left(
G\right) -dim\left( H_{0}\right) =2n_{V}+1=dim_{\mathbb{R}}\left(
R_{V}\right) -1.
\end{array}
\end{equation}
\bigskip

We need to consider the $N=2$ Attractor Eqs.; these are nothing but the
criticality conditions for the $N=2$ BH effective potential \cite{CDF,FK1}
\begin{equation}
V_{BH}\equiv \left| Z\right| ^{2}+G^{i\overline{i}}D_{i}Z\overline{D}_{%
\overline{i}}\overline{Z}  \label{VBH-def}
\end{equation}
in the corresponding special K\"{a}hler geometry \cite{FGK}:
\begin{equation}
\partial _{i}V_{BH}=0\Longleftrightarrow 2\overline{Z}D_{i}Z+iC_{ijk}G^{j%
\overline{j}}G^{k\overline{k}}\overline{D}_{\overline{j}}\overline{Z}%
\overline{D}_{\overline{k}}\overline{Z}=0,\forall i=1,...,n_{V}.
\label{AEs1}
\end{equation}
$C_{ijk}$ is the rank-$3$, completely symmetric, covariantly holomorphic
tensor of special K\"{a}hler geometry, satisfying (see e.g. \cite{CDFVP})
\begin{equation}
\overline{D}_{\overline{l}}C_{ijk}=0,~~~D_{[l}C_{i]jk}=0,  \label{propr-C}
\end{equation}
where the square brackets denote antisymmetrization with respect to the
enclosed indices.

For symmetric special K\"{a}hler manifolds the tensor $C_{ijk}$ is
covariantly constant:\textbf{\ }
\begin{equation}
D_{i}C_{jkl}=0,  \label{CERN1}
\end{equation}
which further implies \cite{GST2,CVP}\textbf{\ }
\begin{equation}
G^{k\overline{k}}G^{r\overline{j}}C_{r(pq}C_{ij)k}\overline{C}_{\overline{k}%
\overline{i}\overline{j}}=\frac{4}{3}G_{\left( q\right| \overline{i}%
}C_{\left| ijp\right) }.  \label{CERN2}
\end{equation}
This equation is simply the $d=4$ version of the \textit{``adjoint identity''%
} satisfied by all (Euclidean) Jordan algebras of degree three that define
the corresponding MESGT's in $d=5$ \cite{GST2,CVP}:
\begin{equation}
d_{r(pq}d_{ij)k}d^{rkl}=\frac{4}{3}\delta _{(q}^{l}d_{ijp)}.
\end{equation}

$Z$ is the $N=2$ \textit{``central charge''} function, whereas $\left\{
D_{i}Z\right\} _{i=1,...,n_{V}}$ is the set of its K\"{a}hler-covariant
holomorphic derivatives, which are nothing but the \textit{``matter charge''%
} functions of the system. Indeed, the sets\footnote{%
We always consider the ``classical'' framework, disregarding the
actual
quantization of the ranges of the electric and magnetic charges $q_{0}$, $%
q_{i}$, $p^{0}$ and $p^{i}$. That is why we consider $\mathbb{R}^{2n_{V}+2}$
rather than the $\left( 2n_{V}+2\right) $-dim. charge lattice $\widehat{%
\Gamma }_{(p,q)}$.} $\left\{ q_{0},q_{i},p^{0},p^{i}\right\} \in \mathbb{R}%
^{2n_{V}+2}$ and $\left\{ Z,D_{i}Z\right\} \in \mathbb{C}^{n_{V}+1}$ (when
evaluated at purely $\left( q,p\right) $-dependent critical values of the
moduli) are two equivalent basis for the charges of the system, and they are
related by a particular set of identities of special K\"{a}hler geometry
\cite{CDF,BFM,AoB-book}. The decomposition (\ref{decomp1}) corresponds to
nothing but the splitting of the sets $\left\{
q_{0},q_{i},p^{0},p^{i}\right\} $ ($\left\{ Z,D_{i}Z\right\} $) of $2n_{V}+2$
($n_{V}+1$) real\ (complex) charges (``charge'' functions) in $q_{0},p^{0}$ (%
$Z$) (related to the graviphoton, and corresponding to $\mathbf{1}_{\mathbb{C%
}}+c.c.$) and in $\left\{ q_{i},p^{i}\right\} $ ($\left\{ D_{i}Z\right\} $)
(related to the $n_{V}$\ vector multiplets, and corresponding to $%
R_{H_{0}}+c.c.$).\medskip

In order to perform the subsequent analysis of orbits, it is convenient to
use ``flat'' $I$-indices by using the (inverse) $n_{V}$-bein $e_{I}^{i}$ of $%
\frac{G}{H_{0}\times U(1)}$:
\begin{equation}
D_{I}Z=e_{I}^{i}D_{i}Z.
\end{equation}
By switching to ``flat'' local $I$-indices, the special K\"{a}hler metric $%
G_{i\overline{j}}$ (assumed to be \textit{regular}, \textit{i.e.} strictly
positive definite everywhere) will become nothing but the Euclidean $n_{V}$%
-dim. metric $\delta _{I\overline{J}}$. Thus, the attractor eqs. (\ref{AEs1}%
) can be ``flattened'' as follows:
\begin{equation}
\partial _{I}V_{BH}=0\Longleftrightarrow 2\overline{Z}D_{I}Z+iC_{IJK}\delta
^{J\overline{J}}\delta ^{K\overline{K}}\overline{D}_{\overline{J}}\overline{Z%
}\overline{D}_{\overline{K}}\overline{Z}=0,\forall I=1,...,n_{V}.
\label{AEs2}
\end{equation}
Note that $C_{IJK}$ becomes an $H_{0}$-invariant tensor \cite{ADFFF}. This
is possible because $C_{ijk}$ in special coordinates is proportional to the
invariant tensor $d_{IJK}$ of the $d=5$ $U$-duality group $G_{5}$. $G_{5}$
and $H_{0}$ correspond to two different real forms of the same Lie algebra
\cite{GST2}.

As it is well known, $\frac{1}{2}$-BPS attractors are given by the following
solution \cite{FGK} of attractor eqs. (\ref{AEs1}) and (\ref{AEs2}):
\begin{equation}
Z\neq 0,\text{ }D_{i}Z=0\Leftrightarrow D_{I}Z=0,\forall i,I=1,...,n_{V}.
\label{BPS}
\end{equation}

Since the ``flattened matter charges'' $D_{I}Z$\ are a vector of
$R_{H_{0}}$, Eq. (\ref{BPS}) directly yields that
$\frac{1}{2}$-BPS solutions are
manifestly $H_{0}$-invariant. In other words, since the $N=2$, $\frac{1}{2}$%
-BPS orbits are of the form $\frac{G}{H_{0}}$, the condition for the $\left(
n_{V}+1\right) $-dim. complex vector $\left( Z,D_{i}Z\right) $ to be $H_{0}$%
-invariant is precisely given by Eq. (\ref{BPS}), defining $N=2$, $\frac{1}{2%
}$-BPS attractor solutions.\smallskip

Thus, as for the $N=8$, $d=4$ attractor solutions (see \textit{e.g.} \cite
{BFGM1} and Refs. therein), also for the $N=2$, $d=4$ $\frac{1}{2}$-BPS case
\textit{the invariance properties of the solutions at the critical point(s)
are given by the maximal compact subgroup (mcs)  of the stabilizer of the corresponding charge orbit, }%
which in the present case is the compact stabilizer itself. Thus, at $N=2$ $%
\frac{1}{2}$-BPS critical points the following enhancement of symmetry
holds:
\begin{equation}
\mathcal{S}\longrightarrow H_{0},  \label{symm-en-BPS}
\end{equation}
where here and below $\mathcal{S}$ denotes the compact symmetry of a generic
orbit of the real symplectic representation $R_{V}$ of the $d=4$ duality
group $G$.\bigskip

However, all the scalar manifolds of $N=2$, $d=4$ symmetric MESGT's have
other species of \textit{regular} critical points $V_{BH}$ (and
correspondingly other classes of \textit{``large''} charge orbits).

Concerning the $N=2$, $d=4$ symmetric MESGT's, the rank-1 sequence $I$ has
one more, non-BPS class of orbits (with vanishing central charge), while all
rank-3 aforementioned cases $II$-$VI$ have two more distinct non-BPS classes
of orbits, one of which with vanishing central charge.

The results about the classes of \textit{``large''} charge orbits of $N=2$, $%
d=4$\ symmetric MESGT's are summarized in Table 3\footnote{%
We should note  that the column on the right of Table 2 of \cite
{FG1} is not fully correct.
\par
Indeed, that column coincides with the central column of Table 3 of
the present paper (by disregarding case $I$ and shifting
$n\rightarrow n-2$ in case $II$), listing the non-BPS, $Z\neq 0$
orbits of $N=2$, $d=4$ symmetric
MESGT's, which are all characterized by a strictly negative quartic $E_{7}$%
-invariant $\mathcal{I}_{4}$. This does not match what is claimed in \cite
{FG1}, where such a column is stated to list the particular class of orbits
with $I_{4}>0$ and eigenvalues of opposite sign in pair.
\par
Actually, the statement of \cite{FG1}\textbf{\ }holds true only for the case
$I$ (which, by shifting $n\rightarrow n-1$, coincides with the last entry of
the column on the right of Table 2 of \cite{FG1}). On the other hand, such a
case is the only one which cannot be obtained from $d=5$ by dimensional
reduction. Moreover, it is the only one not having non-BPS, $Z\neq 0$
orbits, rather it is characterized only by a class of non-BPS orbits with $%
Z=0$ and $\mathcal{I}_{4}>0.$}.

\begin{table}[t]
\begin{center}
\begin{tabular}{|c||c|c|c|}
\hline
& $
\begin{array}{c}
\\
\frac{1}{2}\text{-BPS orbits } \\
~~\mathcal{O}_{\frac{1}{2}-BPS}=\frac{G}{H_{0}} \\
~
\end{array}
$ & $
\begin{array}{c}
\\
\text{non-BPS, }Z\neq 0\text{ orbits} \\
\mathcal{O}_{non-BPS,Z\neq 0}=\frac{G}{\widehat{H}}~ \\
~
\end{array}
$ & $
\begin{array}{c}
\\
\text{non-BPS, }Z=0\text{ orbits} \\
\mathcal{O}_{non-BPS,Z=0}=\frac{G}{\widetilde{H}}~ \\
~
\end{array}
$ \\ \hline\hline
$
\begin{array}{c}
\\
I \\
~
\end{array}
$ & $\frac{SU(1,n+1)}{SU(n+1)}~$ & $-$ & $\frac{SU(1,n+1)}{SU(1,n)}~$ \\
\hline
$
\begin{array}{c}
\\
II \\
~
\end{array}
$ & $SU(1,1)\times \frac{SO(2,2+n)}{SO(2)\times SO(2+n)}~$ & $SU(1,1)\times
\frac{SO(2,2+n)}{SO(1,1)\times SO(1,1+n)}~$ & $SU(1,1)\times\frac{ SO(2,2+n)%
}{SO(2)\times SO(2,n)}$ \\ \hline
$
\begin{array}{c}
\\
III \\
~
\end{array}
$ & $\frac{E_{7(-25)}}{E_{6}}$ & $\frac{E_{7(-25)}}{E_{6(-26)}}$ & $\frac{%
E_{7(-25)}}{E_{6(-14)}}~$ \\ \hline
$
\begin{array}{c}
\\
IV \\
~
\end{array}
$ & $\frac{SO^{\ast }(12)}{SU(6)}~$ & $\frac{SO^{\ast }(12)}{SU^{\ast }(6)}~$
& $\frac{SO^{\ast }(12)}{SU(4,2)}~$ \\ \hline
$
\begin{array}{c}
\\
V \\
~
\end{array}
$ & $\frac{SU(3,3)}{SU(3)\times SU(3)}$ & $\frac{SU(3,3)}{SL(3,\mathbb{C})}$
& $\frac{SU(3,3)}{SU(2,1)\times SU(1,2)}~$ \\ \hline
$
\begin{array}{c}
\\
VI \\
~
\end{array}
$ & $\frac{Sp(6,\mathbb{R})}{SU(3)}$ & $\frac{Sp(6,\mathbb{R})}{SL(3,\mathbb{%
R})}$ & $\frac{Sp(6,\mathbb{R})}{SU(2,1)}$ \\ \hline
\end{tabular}
\end{center}
\caption{\textit{``Large}\textbf{'' orbits of $N=2$, $d=4$ symmetric MESGT's
}}
\end{table}
\def\theequation{2.\arabic{subsection}.\arabic{equation}}
\subsection{\label{N=2-Attractors}Classification of Attractors}

The three classes of orbits in Table 3 correspond to the three distinct
classes of solutions of the $N=2$, $d=4$ Attractor Eqs. (\ref{AEs1}) and (%
\ref{AEs2}). \setcounter{equation}0
\def\theequation{2.2.\arabic{subsubsection}.\arabic{equation}}
\subsubsection{\label{N=2-Attractors-BPS}$\frac{1}{2}$-BPS}

As already mentioned, the class of $\frac{1}{2}$-BPS orbits corresponds to
the solution (\ref{BPS}) determining $N=2$, $\frac{1}{2}$-BPS critical
points of $V_{BH}$. Such a solution yields the following value of the BH
scalar potential at the considered attractor point(s) \cite{FGK}:
\begin{equation}
V_{BH,\frac{1}{2}-BPS}=\left| Z\right| _{\frac{1}{2}-BPS}^{2}+\left[ G^{i%
\overline{i}}D_{i}Z\overline{D}_{\overline{i}}\overline{Z}\right] _{\frac{1}{%
2}-BPS}=\left| Z\right| _{\frac{1}{2}-BPS}^{2}.
\end{equation}
\textit{The overall symmetry group at }$N=2$\textit{\ }$\frac{1}{2}$\textit{%
-BPS critical point(s) is }$H_{0}$\textit{, stabilizer of }$~\mathcal{O}_{%
\frac{1}{2}-BPS}=\frac{G}{H_{0}}$. The \textit{symmetry enhancement} is
given by Eq. (\ref{symm-en-BPS}). For such a class of orbits
\begin{equation}
I_{4,\frac{1}{2}-BPS}=\left| Z\right| _{\frac{1}{2}-BPS}^{4}>0.
\end{equation}
\setcounter{equation}0
\def\theequation{2.2.\arabic{subsubsection}.\arabic{equation}}
\subsubsection{\label{N=2-Attractors-non-BPS}Non-BPS}

The two classes of $N=2$ non-BPS \textit{``large''} charge orbits
respectively correspond to the following solutions of $N=2$ attractor eqs. (%
\ref{AEs1}):
\begin{equation}
\text{non-BPS, }Z\neq 0\text{:}\left\{
\begin{array}{l}
Z\neq 0, \\
\\
D_{i}Z\neq 0\text{ for some }i\in \left\{ 1,...,n_{V}\right\} , \\
\\
I_{4,non-BPS,Z\neq 0}=-\left( \left| Z\right| _{non-BPS,Z\neq 0}^{2}+\left(
G^{i\overline{j}}D_{i}Z\overline{D}_{\overline{j}}\overline{Z}\right)
_{non-BPS,Z\neq 0}\right) ^{2}= \\
\\
=-16\left| Z\right| _{non-BPS,Z\neq 0}^{4}<0;
\end{array}
\right.
\end{equation}
\begin{equation}
\text{non-BPS, }Z=0\text{:}\left\{
\begin{array}{l}
Z=0, \\
\\
D_{i}Z\neq 0\text{ for some }i\in \left\{ 1,...,n_{V}\right\} , \\
\\
I_{4,non-BPS,Z=0}=\left( G^{i\overline{j}}D_{i}Z\overline{D}_{\overline{j}}%
\overline{Z}\right) _{non-BPS,Z=0}^{2}>0.
\end{array}
\right.
\end{equation}

In the treatment given below, we will show how the general solutions of Eqs. (%
\ref{AEs1}), respectively determining the two aforementioned classes of $N=2$
non-BPS extremal BH attractors, can be easily given by using ``flat'' local $%
I$-coordinates in the scalar manifold.

In other words, we will consider the ``flattened'' attractor eqs. (\ref{AEs2}%
), which can be specialized in the \textit{``large''} non-BPS cases as
follows:
\begin{eqnarray}
&&
\begin{array}{l}
\text{non-BPS, }Z\neq 0\text{:~~~}2\overline{Z}D_{I}Z=-iC_{IJK}\delta ^{J%
\overline{J}}\delta ^{K\overline{K}}\overline{D}_{\overline{J}}\overline{Z}%
\overline{D}_{\overline{K}}\overline{Z};
\end{array}
\notag \\
&&  \label{AEs-non-BPS1-flat} \\
&&
\begin{array}{l}
\text{non-BPS, }Z=0\text{:~~~}C_{IJK}\delta ^{J\overline{J}}\delta ^{K%
\overline{K}}\overline{D}_{\overline{J}}\overline{Z}\overline{D}_{\overline{K%
}}\overline{Z}=0.
\end{array}
\notag \\
&&  \label{AEs-non-BPS2-flat}
\end{eqnarray}
Thus, by respectively denoting with $\widehat{H}$ ($\widetilde{H}$) the
stabilizer of the $N=2$, non-BPS, $Z\neq 0$ ($Z=0$) classes of orbits listed
in Table 3, our claim is the following: \textit{the general solution of Eqs.
}(\ref{AEs-non-BPS1-flat})\textit{\ (}(\ref{AEs-non-BPS2-flat})\textit{) is
obtained by retaining a complex charge vector }$\left( Z,D_{I}Z\right) $%
\textit{\ which is invariant under }$\widehat{h}$\textit{\ (}$\frac{%
\widetilde{h}}{U(1)}$\textit{), where }$\widehat{h}$\textit{\ (}$\widetilde{h%
}$\textit{) is the mcs}\footnote{%
Indeed, while $H_{0}$ is a proper compact subgroup of $G$, the groups $%
\widehat{H}$, $\widetilde{H}$ are real (non-compact) forms of $H_{0}$, as it
can be seen from Table 3 (see also\textbf{\ }\cite{Gilmore,Helgason}).
Therefore in general they admit a \textit{mcs} $\widehat{h}$, $\widetilde{h}$%
, which in turn is a (non-maximal) compact subgroup of $G$ and a proper
compact subgroup of $H_{0}$.
\par
It is interesting to notice that in all cases (listed in Table 3) $G$ always
admits only 2 real (non-compact) forms $\widehat{H}$, $\widetilde{H}$ of $%
H_{0}$ as proper subgroups (consistent with the required dimension of
orbits). The inclusion of $\widehat{H}$, $\widetilde{H}$ in $G$ is such that
in all cases $\widehat{H}\times SO(1,1)$ and $\widetilde{H}\times U(1)$ are
different maximal non-compact subgroups of $G$.}\textit{\ of }$\widehat{H}$%
\textit{\ (}$\widetilde{H}$\textit{)}.

As a consequence, \textit{the overall symmetry group of the }$N=2$\textit{,
non-BPS, }$Z\neq 0$\textit{\ (}$Z=0$\textit{) critical point(s) is }$%
\widehat{h}$\textit{\ (}$\frac{\widetilde{h}}{U(1)}$\textit{). }Thus, at $%
N=2 $, non-BPS, $Z\neq 0$\ ($Z=0$) critical point(s) the following \textit{%
enhancement of symmetry} holds:
\begin{equation}
\begin{array}{l}
N=2\text{, non-BPS, }Z\neq 0:\mathcal{S}\longrightarrow \widehat{h}\
=mcs\left( \widehat{H}\right) ; \\
\\
N=2\text{, non-BPS, }Z=0:\mathcal{S}\longrightarrow \ \frac{\widetilde{h}}{%
U(1)}=\frac{mcs\left( \widetilde{H}\right) }{U(1)}.
\end{array}
\label{symm-en-non-BPS}
\end{equation}

It is worth remarking that the non-compact group $\widehat{H}$ stabilizing
the non-BPS, $Z\neq 0$ class of orbits of $N=2$, $d=4$ symmetric MESGT's,
beside being a real (non-compact) form of $H_{0}$, \ is isomorphic to the
duality group $G_{5}$ of $N=2$, $d=5$ symmetric MESGT's\footnote{%
Such a feature is missing in the $N=2$, $d=4$ symmetric MESGT's whose scalar
manifolds belong to the sequence $I$ given by Eq. (\ref{quadr}), simply
because such theories do not have a class of non-BPS, $Z\neq 0$ orbits.}.

Since the scalar manifolds of $N=2$, $d=5$ symmetric MESGT's are endowed
with a real special geometry \cite{GST1,GST2,GST3}, the complex
representation $R_{H_{0}}$ of $H_{0}$ decomposes in a pair of irreducible
real representations $\left( R_{\widehat{h}}+\mathbf{1}\right) _{\mathbb{R}}$%
's of $\widehat{h}=mcs\left( \widehat{H}\right) \varsubsetneq $ $H_{0}$ (see
Subsubsect. \ref{N=2-Attractors-non-BPS-1}, and in particular Eq. (\ref{KR}%
)). As we will see below, such a fact crucially distinguishes the
non-BPS, $Z\neq 0$ and $Z=0$ cases.

The stabilizers (and the corresponding \textit{mcs}'s) of the non-BPS, $%
Z\neq 0$ and $Z=0$ classes of orbits of $N=2$, $d=4$ symmetric MESGT's are
given in Table 4.

\begin{table}[t]
\begin{center}
\begin{tabular}{|c||c|c|c|c|c|}
\hline
& $
\begin{array}{c}
\\
H_{0} \\
~
\end{array}
$ & $
\begin{array}{c}
\\
\widehat{H} \\
~
\end{array}
$ & $
\begin{array}{c}
\\
\widetilde{H} \\
~
\end{array}
$ & $
\begin{array}{c}
\\
\widehat{h}\equiv mcs\left( \widehat{H}\right) \\
~
\end{array}
$ & $
\begin{array}{c}
\\
\widetilde{h}^{\prime }\equiv \frac{mcs\left( \widetilde{H}\right) }{U(1)}
\\
~
\end{array}
$ \\ \hline\hline
$I$ & $
\begin{array}{c}
\\
SU(n+1) \\
~
\end{array}
$ & $-$ & $SU(1,n)$ & $-$ & $SU(n)$ \\ \hline
$II$ & $
\begin{array}{c}
SO(2) \\
\times \\
SO(2+n)
\end{array}
$ & $
\begin{array}{c}
SO(1,1) \\
\times \\
SO(1,1+n)
\end{array}
$ & $
\begin{array}{c}
SO(2) \\
\times \\
SO(2,n)
\end{array}
$ & $SO(1+n)$ & $
\begin{array}{c}
SO(2) \\
\times \\
SO(n)
\end{array}
$ \\ \hline
$III$ & $
\begin{array}{c}
\\
E_{6}\equiv E_{6(-78)} \\
~
\end{array}
$ & $E_{6(-26)}$ & $E_{6(-14)}$ & $F_{4}\equiv F_{4(-52)}$ & $SO(10)$ \\
\hline
$IV$ & $SU(6)$ & $SU^{\ast }(6)$ & $SU(4,2)$ & $USp(6)$ & $
\begin{array}{c}
SU(4) \\
\times \\
SU(2)
\end{array}
$ \\ \hline
$V$ & $SU(3)\times SU(3)$ & $SL(3,\mathbb{C})$ & $
\begin{array}{c}
SU(2,1) \\
\times \\
SU(1,2)
\end{array}
$ & $SU(3)$ & $
\begin{array}{c}
SU(2) \\
\times \\
SU(2)\times U(1)
\end{array}
$ \\ \hline
$VI$ & $
\begin{array}{c}
\\
SU(3) \\
~
\end{array}
$ & $SL(3,\mathbb{R})$ & $SU(2,1)$ & $SO(3)$ & $SU(2)$ \\ \hline
\end{tabular}
\end{center}
\caption{\textbf{Stabilizers and corresponding maximal compact subgroups of
the \textit{``large''} classes of orbits of }$N=2$\textbf{, }$d=4$\textbf{\
symmetric MESGT's}. {$\widehat{H}$ and $\widetilde{H}$ are real
(non-compact) forms of $H_{0}$, the stabilizer of $\frac{1}{2}$-BPS orbits.}}
\end{table}

\paragraph{\label{N=2-Attractors-non-BPS-1}Non-BPS, $Z\neq 0$}

Let us start by considering the class of non-BPS, $Z\neq 0$ orbits of $N=2$,
$d=4$ symmetric MESGT's.

As mentioned, the ``flattened matter charges'' $D_{I}Z$ are a vector of $%
R_{H_{0}}$. In general, $R_{H_{0}}$ decomposes under the \textit{mcs} $%
\widehat{h}\subset \widehat{H}$ as follows:
\begin{equation}
R_{H_{0}}\longrightarrow \left( R_{\widehat{h}}+\mathbf{1}\right) _{\mathbb{C%
}}\mathbf{,}  \label{KR}
\end{equation}
where the r.h.s. is made of the complex singlet representation of $\widehat{h%
}$ and by another non-singlet real representation of $\widehat{h}$, denoted
above with $R_{\widehat{h}}$. As previously mentioned, despite being
complex, $\left( R_{\widehat{h}}+\mathbf{1}\right) _{\mathbb{C}}$ is not
charged with respect to $U(1)$ symmetry because, due to the 5-dimensional
origin of the non-compact stabilizer $\widehat{H}$ whose \textit{mcs} is $%
\widehat{h}$, actually $\left( R_{\widehat{h}}+\mathbf{1}\right) _{\mathbb{C}%
}$ is nothing but the complexification of its real counterpart$\left( R_{%
\widehat{h}}+\mathbf{1}\right) _{\mathbb{R}}$. The decomposition (\ref{KR}%
)yields the following splitting of ``flattened matter charges'':
\begin{equation}
D_{I}Z\longrightarrow \left( D_{\widehat{I}}Z,D_{\widehat{I}_{0}}Z\right) ,
\end{equation}
where $\widehat{I}$ are the indices along the representation $R_{\widehat{h}%
} $, and $\widehat{I}_{0}$ is the $\widehat{h}$-singlet index.

By considering the related attractor eqs., it should be noticed that the
rank-3 symmetric tensor $C_{IJK}$ in Eqs.\textit{\ }(\ref{AEs-non-BPS1-flat}%
) corresponds to a cubic $H_{0}$-invariant coupling $\left( R_{H_{0}}\right)
^{3}$. By decomposing $\left( R_{H_{0}}\right) ^{3}$ in terms of
representations of $\widehat{h}$, one finds
\begin{equation}
\left( R_{H_{0}}\right) ^{3}\longrightarrow \left( R_{\widehat{h}}\right)
^{3}+\left( R_{\widehat{h}}\right) ^{2}\mathbf{1}_{\mathbb{C}}+\left(
\mathbf{1}_{\mathbb{C}}\right) ^{3}\mathbf{.}  \label{decomp-non-BPS-Z<>0}
\end{equation}
Notice that a term $R_{\widehat{h}}\left( \mathbf{1}_{\mathbb{C}}\right)
^{2} $ cannot be in such a representation decomposition, since it is not $%
\widehat{h}$-invariant, and thus not $H_{0}$-invariant. This implies that
components of the form $C_{\widehat{I}\widehat{I}_{0}\widehat{I}_{0}}$
cannot exist. Also, a term like $\left( \mathbf{1}_{\mathbb{C}}\right) ^{3}$
can appear in the r.h.s. of the decomposition (\ref{AEs-non-BPS1-flat})
since as we said the $\widehat{h}$-singlet $\mathbf{1}_{\mathbb{C}}$,
despite being complex, is \textit{not} $U(1)$-charged.

It is then immediate to conclude that the solution of $N=2$, $d=4$ non-BPS, $%
Z\neq 0$ extremal BH attractor eqs. in ``flat'' indices (\ref
{AEs-non-BPS1-flat}) corresponds to keeping the \textit{``flattened matter charges''} $%
D_{I}Z$ $\widehat{h}$-invariant. By virtue of decomposition (\ref
{decomp-non-BPS-Z<>0}), this is obtained by putting
\begin{equation}
D_{\widehat{I}}Z=0,D_{\widehat{I}_{0}}Z\neq 0,  \label{sol-non-BPS-Z<>0}
\end{equation}
i.e. by putting all ``flattened matter charges'' to zero, except
the one along the $\widehat{h}$-singlet (and thus
$\widehat{h}$-invariant) direction in scalar manifold. By
substituting the solution (\ref{sol-non-BPS-Z<>0}) in Eqs.
(\ref{AEs-non-BPS1-flat}), one obtains
\begin{gather}
2\overline{Z}D_{\widehat{I}_{0}}Z=-iC_{\widehat{I}_{0}\widehat{I}_{0}%
\widehat{I}_{0}}\left( \overline{D}_{\overline{\widehat{I}_{0}}}\overline{Z}%
\right) ^{2}\overset{Z\neq 0}{\Leftrightarrow }D_{\widehat{I}_{0}}Z=-\frac{i%
}{2}\frac{C_{\widehat{I}_{0}\widehat{I}_{0}\widehat{I}_{0}}}{\overline{Z}}%
\left( \overline{D}_{\overline{\widehat{I}_{0}}}\overline{Z}\right) ^{2} \\
\Downarrow  \notag \\
\left| D_{\widehat{I}_{0}}Z\right| ^{2}\left( 1-\frac{1}{4}\frac{\left| C_{%
\widehat{I}_{0}\widehat{I}_{0}\widehat{I}_{0}}\right| ^{2}}{\left| Z\right|
^{2}}\left| D_{\widehat{I}_{0}}Z\right| ^{2}\right) =0  \notag \\
\Updownarrow  \notag \\
\left| D_{\widehat{I}_{0}}Z\right| ^{2}=4\frac{\left| Z\right| ^{2}}{\left|
C_{\widehat{I}_{0}\widehat{I}_{0}\widehat{I}_{0}}\right| ^{2}};
\label{sol-non-BPS-Z<>0-2}
\end{gather}
this is nothing but the general criticality condition of $V_{BH}$ for the
1-modulus case in the locally ``flat'' coordinate $\widehat{I}_{0}$, which
in this case corresponds to the $\widehat{h}$-singlet direction in the
scalar manifold. Such a case has been thoroughly studied in non-flat $i$%
-coordinates in \cite{BFM}.\smallskip

All $N=2$, $d=4$ symmetric MESGT's (disregarding the sequence $I$ having $%
C_{ijk}=0$) have a cubic prepotential ($F=\frac{1}{3!}d_{ijk}t^{i}t^{j}t^{k}$
in special coordinates), and thus in special coordinates it holds that $%
C_{ijk}=e^{K}d_{ijk}$, with $K$ and $d_{ijk}$ respectively denoting the
K\"{a}hler potential and the completely symmetric rank-3 constant tensor
that is determined by the norm form of the underlying Jordan algebra of
degree three \cite{GST2}. In the cubic $n_{V}=1$-modulus case, by using Eq. (%
\ref{CERN2}) it follows that
\begin{equation}
\left( G^{1\overline{_{s}1}_{s}}\right) ^{3}\left|
C_{1_{s}1_{s}1_{s}}\right| ^{2}=\left| C_{1_{f}1_{f}1_{f}}\right| ^{2}=\frac{%
4}{3},  \label{ress1}
\end{equation}
where the subscripts ``$s$'' and ``$f$'' respectively stand for \textit{%
``special''} and \textit{``flat''}, denoting the kind of coordinate system
being considered. By substituting Eq. (\ref{ress1}) in Eq. (\ref
{sol-non-BPS-Z<>0-2}) one obtains the result
\begin{equation}
\left| D_{\widehat{I}_{0}}Z\right| ^{2}=3\left| Z\right| ^{2}.  \label{ress2}
\end{equation}
Another way of proving Eq. (\ref{ress2})\ is by computing the quartic
invariant along the $\widehat{h}$-singlet direction, then yielding
\begin{equation}
I_{4,non-BPS,Z\neq 0}=-16\left| Z\right| _{non-BPS,Z\neq 0}^{2}.
\end{equation}

The considered solution (\ref{sol-non-BPS-Z<>0})-(\ref{sol-non-BPS-Z<>0-2}),
(\ref{ress2}) is the $N=2$ analogue of the $N=8$, $d=4$ non-BPS \textit{%
``large''} solution discussed in \cite{FKlast}, and it yields the following
value of the BH scalar potential at the considered attractor point(s) \cite
{BFM,FKlast}:
\begin{equation}
V_{BH,non-BPS,Z\neq 0}=4\left| Z\right| _{non-BPS,Z\neq 0}^{2}.
\label{ress3}
\end{equation}
Once again, as for the non-BPS $N=8$ \textit{``large''} solutions \label%
{FKlast,BFGM1}, we find the extra factor $4$.

From the above considerations, \textit{the overall symmetry group at }$N=2$%
\textit{\ non-BPS, }$Z\neq 0$\textit{\ critical point(s) is }$\widehat{h}$%
\textit{, mcs of the non-compact stabilizer }$\widehat{H}$\textit{\ of }$%
\mathcal{O}_{non-BPS,Z\neq 0}$\textit{.}

\paragraph{\label{N=2-Attractors-non-BPS-2}Non-BPS, $Z=0$}

Let us now move to consider the other class of non-BPS orbits of $N=2$, $d=4$
symmetric MESGT's.

It has $Z=0$ and it was not considered in \cite{FG1} (see also
Footnote 4). We will show that the solution of the $N=2$, $d=4$,
non-BPS, $Z=0$ extremal BH attractor eqs.
(\ref{AEs-non-BPS2-flat}) are the ``flattened matter charges''
$D_{I}Z$ which are invariant under $\frac{\widetilde{h}}{U(1)}$,
where $\widetilde{h}$ is the \textit{mcs} of $\widetilde{H}$, the
stabilizer of the class
$\mathcal{O}_{non-BPS,Z=0}=\frac{G}{\widetilde{H}}$.

Differently from the non-BPS, $Z\neq 0$ case, in the considered non-BPS, $%
Z=0 $ case there is always a $U(1)$ symmetry acting, since the scalar
manifolds of $N=2$, $d=4$ symmetric MESGT's \textit{all} have the group $%
\widetilde{h}$ of the form
\begin{equation}
\widetilde{h}=\widetilde{h}^{\prime }\times U(1),\text{ \ \ }\widetilde{h}%
^{\prime }\equiv \frac{\widetilde{h}}{U(1)}.  \label{gen-struct}
\end{equation}

The compact subgroups $\widetilde{h}^{\prime }$ for all $N=2$, $d=4$
symmetric MESGT's are listed in Table 4. In the case at hand, we thence have
to consider the decomposition of the previously introduced complex
representation $R_{H_{0}}$ under the compact subgroup $\widetilde{h}^{\prime
}\varsubsetneq H_{0}$. In general, $R_{H_{0}}$ decomposes under $\widetilde{h%
}^{\prime }\varsubsetneq \widetilde{H}$ as follows:
\begin{equation}
R_{H_{0}}\longrightarrow \left( \mathcal{W}_{\widetilde{h}^{\prime }}+%
\mathcal{Y}_{\widetilde{h}^{\prime }}+\mathbf{1}\right) _{\mathbb{C}}\mathbf{%
,}  \label{KR-KR-KR}
\end{equation}
where in the r.h.s. the complex singlet representation of $\widetilde{h}%
^{\prime }$ and two complex non-singlet representations $\mathcal{W}_{%
\widetilde{h}^{\prime }}$ and $\mathcal{Y}_{\widetilde{h}^{\prime }}$ of $%
\widetilde{h}^{\prime }$ appear. In general, $\mathcal{W}_{\widetilde{h}%
^{\prime }}$, $\mathcal{Y}_{\widetilde{h}^{\prime }}$ and $\mathbf{1}_{%
\mathbb{C}}$ are charged (and thus not invariant) with respect to
the $U(1)$ explicit factor appearing in (\ref{gen-struct}). The
decomposition (\ref {KR-KR-KR}) yields the following splitting of
``flattened matter charges'':
\begin{equation}
D_{I}Z\longrightarrow \left( D_{\widetilde{I}_{\mathcal{W}}^{\prime }}Z,D_{%
\widetilde{I}_{\mathcal{Y}}^{\prime }}Z,D_{\widetilde{I}_{0}^{\prime
}}Z\right) ,
\end{equation}
where $\widetilde{I}_{\mathcal{W}}^{\prime }$ and $\widetilde{I}_{\mathcal{Y}%
}^{\prime }$ respectively denote the indices along the complex
representations $\mathcal{W}_{\widetilde{h}^{\prime }}$ and $\mathcal{Y}_{%
\widetilde{h}^{\prime }}$, and $\widetilde{I}_{0}^{\prime }$ is the $%
\widetilde{h}^{\prime }$-singlet index.

Once again, the related $N=2$, $d=4$ non-BPS, $Z=0$ extremal BH attractor
eqs. (\ref{AEs-non-BPS2-flat}) contain the rank-3 symmetric tensor $C_{IJK}$%
, corresponding to a cubic $H_{0}$-invariant coupling $\left(
R_{H_{0}}\right) ^{3}$. The decomposition of $\left( R_{H_{0}}\right) ^{3}$
in terms of representations of $\widetilde{h}^{\prime }$ yields
\begin{equation}
\left( R_{H_{0}}\right) ^{3}\longrightarrow \left( \mathcal{W}_{\widetilde{h}%
^{\prime }}\right) ^{2}\mathcal{Y}_{\widetilde{h}^{\prime }}+\left( \mathcal{%
Y}_{\widetilde{h}^{\prime }}\right) ^{2}\mathbf{1}_{\mathbb{C}}\mathbf{.}
\label{decomp-non-BPS-Z=0}
\end{equation}
When decomposed under $\widetilde{h}^{\prime }$, $\left( R_{H_{0}}\right)
^{3}$ must be nevertheless $\widetilde{h}$-invariant, and therefore, beside
the $\widetilde{h}^{\prime }$-invariance, one has to consider the invariance
under the $U(1)$ factor, too. Thus, terms of the form $\left( \mathcal{W}_{%
\widetilde{h}^{\prime }}\right) ^{3}$, $\left( \mathcal{Y}_{\widetilde{h}%
^{\prime }}\right) ^{3}$, $\mathcal{W}_{\widetilde{h}^{\prime }}\left(
\mathbf{1}_{\mathbb{C}}\right) ^{2}$, $\mathcal{Y}_{\widetilde{h}^{\prime
}}\left( \mathbf{1}_{\mathbb{C}}\right) ^{2}$ and $\left( \mathbf{1}_{%
\mathbb{C}}\right) ^{3}$ cannot exist in the $\widetilde{h}$-invariant
r.h.s. of decomposition (\ref{decomp-non-BPS-Z=0}).

Notice also that the structure of the decomposition (\ref{decomp-non-BPS-Z=0}%
) implies that components of the cubic coupling of the form $C_{\widetilde{I}%
_{\mathcal{W}}^{\prime }\widetilde{I}_{0}^{\prime }\widetilde{I}_{0}^{\prime
}}$, $C_{\widetilde{I}_{\mathcal{Y}}^{\prime }\widetilde{I}_{0}^{\prime }%
\widetilde{I}_{0}^{\prime }}$ and $C_{\widetilde{I}_{0}^{\prime }\widetilde{I%
}_{0}^{\prime }\widetilde{I}_{0}^{\prime }}$ cannot exist. For such a
reason, it is immediate to conclude that the solution of $N=2$, $d=4$
non-BPS, $Z=0$ extremal BH attractor eqs. in ``flat'' indices (\ref
{AEs-non-BPS2-flat}) corresponds to keep the ``flattened matter charges'' $%
D_{I}Z$ $\widetilde{h}^{\prime }$-invariant. By virtue of decomposition (\ref
{decomp-non-BPS-Z=0}), this is obtained by putting
\begin{equation}
D_{\widetilde{I}_{\mathcal{W}}^{\prime }}Z=0=D_{\widetilde{I}_{\mathcal{Y}%
}^{\prime }}Z,~~~~D_{\widetilde{I}_{0}^{\prime }}Z\neq 0,
\label{sol-non-BPS-Z=0}
\end{equation}
\textit{i.e.} by putting all ``flattened matter charges'' to zero,
except the
one along the $\widetilde{h}^{\prime }$-singlet (and thus $\widetilde{h}%
^{\prime }$-invariant, but not $U(1)$-invariant and therefore not $%
\widetilde{h}$-invariant) direction in the scalar manifold.\medskip

The considered solution (\ref{sol-non-BPS-Z=0}) does not have any analogue
in $N=8$, $d=4$ supergravity, and it yields the following value of the BH
scalar potential at the considered attractor point(s):
\begin{eqnarray}
V_{BH,non-BPS,Z=0} &=&\left| Z\right| _{non-BPS,Z=0}^{2}+\left[ G^{i%
\overline{i}}D_{i}Z\overline{D}_{\overline{i}}\overline{Z}\right]
_{non-BPS,Z=0}=  \notag \\
&&  \notag \\
&=&\left| D_{\widetilde{I}_{0}^{\prime }}Z\right| _{non-BPS,Z=0}^{2}.
\end{eqnarray}
It is here worth remarking that in the $stu$ model it can be explicitly
computed that \cite{BMOS-1,stu-unveiled}
\begin{equation}
V_{BH,non-BPS,Z=0}=\left| D_{\widetilde{I}_{0}^{\prime }}Z\right|
_{non-BPS,Z=0}^{2}=\left| Z\right| _{\frac{1}{2}-BPS}^{2}=V_{BH,\frac{1}{2}%
-BPS}.  \label{stu-stu}
\end{equation}

From above considerations, \textit{the overall symmetry group at }$N=2$%
\textit{\ non-BPS, }$Z=0$\textit{\ critical point(s) is }$\widetilde{h}%
^{\prime }=\frac{\widetilde{h}}{U(1)}$\textit{, }$\widetilde{h}$\textit{\
being the mcs of the non-compact stabilizer }$\widetilde{H}$\textit{\ of }$%
\mathcal{O}_{non-BPS,Z=0}$\textit{.\medskip }

The general analysis carried out above holds for all $N=2$, $d=4$ symmetric
\textit{magic} MESGT's, namely for the irreducible cases $III$-$VI$ listed
in Tables 2 and 3. The cases of irreducible sequence $I$ and of generic
Jordan family $II$ deserve suitable, slightly different treatments,
respectively given in Appendices I and II of \cite{BFGM1}.
\setcounter{equation}0
\def\theequation{2.\arabic{subsection}.\arabic{equation}}
\subsection{\label{N=2-Spectra}Critical Spectra and Massless Hessian Modes
of $V_{BH}$}

The effective BH potential $V_{BH}$ gives different masses to the different
BPS-phases of the considered symmetric $N=2$, $d=4$ MESGT's. The fundamental
object to be considered in such a framework is the moduli-dependent $%
2n_{V}\times 2n_{V}$ Hessian matrix of $V_{BH}$, which in complex basis reads%
\footnote{%
The reported formul{\ae } for $\mathcal{M}_{ij}$ and $\mathcal{N}_{i%
\overline{j}}$ hold for any special K\"{a}hler manifold. In the symmetric
case formula (\ref{M}) gets simplified using Eq. (\ref{CERN1}).} \cite{BFM}
\begin{eqnarray}
&&
\begin{array}{c}
\mathbf{H}^{V_{BH}}\equiv \left(
\begin{array}{ccc}
D_{i}D_{j}V_{BH} &  & D_{i}\overline{D}_{\overline{j}}V_{BH} \\
&  &  \\
D_{j}\overline{D}_{\overline{i}}V_{BH} &  & \overline{D}_{\overline{i}}%
\overline{D}_{\overline{j}}V_{BH}
\end{array}
\right) \equiv \left(
\begin{array}{ccc}
\mathcal{M}_{ij} &  & \mathcal{N}_{i\overline{j}} \\
&  &  \\
\overline{\mathcal{N}}_{j\overline{i}} &  & \overline{\mathcal{M}}_{%
\overline{i}\overline{j}}
\end{array}
\right) ;
\end{array}
\\
&&  \notag \\
&&  \notag \\
&&
\begin{array}{l}
\mathcal{M}_{ij}\equiv D_{i}D_{j}V_{BH}=D_{j}D_{i}V_{BH}= \\
\\
=4i\overline{Z}C_{ijk}G^{k\overline{k}}\overline{D}_{\overline{k}}\overline{Z%
}+iG^{k\overline{k}}G^{l\overline{l}}\left( D_{j}C_{ikl}\right) \overline{D}%
_{\overline{k}}\overline{Z}\overline{D}_{\overline{l}}\overline{Z};
\end{array}
\label{M} \\
&&  \notag \\
&&  \notag \\
&&
\begin{array}{l}
\mathcal{N}_{i\overline{j}}\equiv D_{i}\overline{D}_{\overline{j}}V_{BH}=%
\overline{D}_{\overline{j}}D_{i}V_{BH}= \\
\\
=2\left[ G_{i\overline{j}}\left| Z\right| ^{2}+D_{i}Z\overline{D}_{\overline{%
j}}\overline{Z}+G^{l\overline{n}}G^{k\overline{k}}G^{m\overline{m}}C_{ikl}%
\overline{C}_{\overline{j}\overline{m}\overline{n}}\overline{D}_{\overline{k}%
}\overline{Z}D_{m}Z\right] ;
\end{array}
\label{N} \\
&&  \notag \\
&&
\begin{array}{l}
\mathcal{M}^{T}=\mathcal{M},\mathcal{N}^{\dag }=\mathcal{N}.
\end{array}
\end{eqnarray}
By analyzing $\mathbf{H}^{V_{BH}}$ at critical points of $V_{BH}$, it is
possible to formulate general conclusions about the mass spectrum of the
corresponding extremal BH solutions with finite, non-vanishing entropy,
\textit{i.e.} about the mass spectrum along the related classes of \textit{%
``large''} charge orbits of the symplectic real representation $R_{V}$ of
the $d=4$ duality group $G$.\smallskip

Let us start by remarking that, due to its very definition (\ref{VBH-def}),
the $N=2$ effective BH potential $V_{BH}$ is positive for any (not
necessarily strictly) positive definite metric $G_{i\overline{i}}$ of the
scalar manifold. Consequently, the \textit{stable} critical points (i.e. the
\textit{attractors} in a strict sense) will necessarily be minima of such a
potential. As already pointed out above and as done also in \cite
{BFM,AoB-book}, the geometry of the scalar manifold is usually assumed to be
\textit{regular, }i.e. endowed with a metric tensor $G_{i\overline{j}}$
being strictly positive definite everywhere.
\def\theequation{2.3.\arabic{subsubsection}.\arabic{equation}}
\subsubsection{\label{BPS-Spectra}$\frac{1}{2}$-BPS}

It is now well known that \textit{regular} special K\"{a}hler geometry
implies that \textit{all} $N=2$ $\frac{1}{2}$-BPS critical points of \textit{%
all} $N=2$, $d=4$ MESGT's are stable, and therefore they are attractors in a
strict sense. Indeed, the Hessian matrix $\mathbf{H}_{\frac{1}{2}%
-BPS}^{V_{BH}}$ evaluated at such points is strictly positive definite \cite
{FGK}:
\begin{equation}
\begin{array}{l}
\mathcal{M}_{ij,\frac{1}{2}-BPS}=0, \\
\\
\mathcal{N}_{i\overline{j},\frac{1}{2}-BPS}=2\left. G_{i\overline{j}}\right|
_{\frac{1}{2}-BPS}\left| Z\right| _{\frac{1}{2}-BPS}^{2}>0,
\end{array}
\label{BPS-crit}
\end{equation}
where the notation ``$>0$'' is clearly understood as strict positive
definiteness of the quadratic form related to the square matrix being
considered. Notice that the Hermiticity and strict positive definiteness of $%
\mathbf{H}_{\frac{1}{2}-BPS}^{V_{BH}}$ are respectively due to the
Hermiticity and strict positive definiteness of the K\"{a}hler metric $G_{i%
\overline{j}}$ of the scalar manifold.

By switching from the non-flat $i$-coordinates to the ``flat'' local $I$%
-coordinates by using the (inverse) Vielbein $e_{I}^{i}$ of the scalar
manifold, Eqs. (\ref{BPS-crit}) can be rewritten as
\begin{equation}
\begin{array}{l}
\mathcal{M}_{IJ,\frac{1}{2}-BPS}=0, \\
\\
\mathcal{N}_{I\overline{J},\frac{1}{2}-BPS}=2\delta _{I\overline{J}}\left|
Z\right| _{\frac{1}{2}-BPS}^{2}>0.
\end{array}
\label{BPS-crit-2}
\end{equation}
Thus, one obtains that in \textit{all} $N=2$, $d=4$ MESGT's the $\frac{1}{2}$%
-BPS mass spectrum in ``flat'' coordinates is \textit{monochromatic}, i.e.
that all ``particles'' (i.e. the ``modes'' related to the degrees of freedom
described by the ``flat'' local $I$-coordinates) acquire \textit{the same}
mass at $\frac{1}{2}$-BPS critical points of $V_{BH}$.
\def\theequation{2.3.\arabic{subsubsection}.\arabic{equation}}
\subsubsection{\label{Non-BPS-Z<>0-Spectra}Non-BPS, $Z\neq 0$}

In this case the result of \cite{TT} should apply, namely the Hessian matrix
$\mathbf{H}_{non-BPS,Z\neq 0}^{V_{BH}}$ should have $n_{V}+1$ strictly
positive and $n_{V}-1$ vanishing real eigenvalues.

By recalling the analysis performed in Sect. \ref{N=2-Attractors}, it is
thence clear that such massive and massless non-BPS, $Z\neq 0$ ``modes'' fit
distinct real representations of $\widehat{h}=mcs\left( \widehat{H}\right) $%
, where $\widehat{H}$ is the non-compact stabilizer of the class $\mathcal{O}%
_{non-BPS,Z\neq 0}=\frac{G}{\widehat{H}}$ of non-BPS, $Z\neq 0$ \textit{%
``large''} charge orbits.

This is perfectly consistent with the decomposition (\ref{KR}) of the
complex representation $R_{H_{0}}$ ($dim_{\mathbb{R}}R_{H_{0}}=2n_{V}$) of $%
H_{0}$ in terms of representations of $\widehat{h}$:
\begin{equation}
R_{H_{0}}\longrightarrow \left( R_{\widehat{h}}+\mathbf{1}\right) _{\mathbb{C%
}}=\left( R_{\widehat{h}}+\mathbf{1}+R_{\widehat{h}}+\mathbf{1}\right) _{%
\mathbb{R}}\mathbf{,}\text{ \ }dim_{\mathbb{R}}\left( R_{\widehat{h}}\right)
_{\mathbb{R}}=n_{V}-1.  \label{KR-KR}
\end{equation}
As yielded by the treatment given in Subsubsect. \ref
{N=2-Attractors-non-BPS-1}, the notation ``$\left( R_{\widehat{h}}+\mathbf{1}%
\right) _{\mathbb{C}}=\left( R_{\widehat{h}}+\mathbf{1}+R_{\widehat{h}}+%
\mathbf{1}\right) _{\mathbb{R}}$'' denotes nothing but the
decomplexification of $\left( R_{\widehat{h}}+\mathbf{1}\right) _{\mathbb{C}%
} $, which is actually composed by a pair of real irreducible
representations $\left( R_{\widehat{h}}+\mathbf{1}\right) _{\mathbb{R}}$ of $%
\widehat{h}$.

Therefore, the result of \cite{TT} can be understood in terms of real
representations of the \textit{mcs} of the non-compact stabilizer of $%
\mathcal{O}_{non-BPS,Z\neq 0}$: the $n_{V}-1$ massless non-BPS, $Z\neq 0$
``modes'' are in one of the two real $R_{\widehat{h}}$'s of $\widehat{h}$ in
the r.h.s. of Eq. (\ref{KR-KR}), say the first one, whereas the $n_{V}+1$
massive non-BPS, $Z\neq 0$ ``modes'' are split in the remaining real $R_{%
\widehat{h}}$ of $\widehat{h}$ and in the two real $\widehat{h}$-singlets.
The resulting interpretation of the decomposition (\ref{KR-KR}) is
\begin{equation}
R_{H_{0}}\longrightarrow \left(
\begin{array}{c}
\left( R_{\widehat{h}}\right) _{\mathbb{R}} \\
\\
n_{V}-1~~\text{\textit{massless} }
\end{array}
\right) +\left(
\begin{array}{c}
\left( R_{\widehat{h}}\right) _{\mathbb{R}}+\mathbf{1}_{\mathbb{R}}+\mathbf{1%
}_{\mathbb{R}} \\
\\
n_{V}+1~~\text{\textit{massive}}
\end{array}
\right) .
\end{equation}
It is interesting to notice once again that there is no $U(1)$ symmetry
relating the two real $R_{\widehat{h}}$'s (and thus potentially relating the
splitting of ``modes'' along $\mathcal{O}_{non-BPS,Z\neq 0}$), since in
\textit{all} symmetric $N=2$, $d=4$ MESGT's $\widehat{h}$ \textit{never}
contains an explicit factor $U(1)$ (as instead it \textit{always} happens
for $\widetilde{h}$!); this can be related to the fact that the non-compact
stabilizer is $\widehat{H}$ whose \textit{mcs} is $\widehat{h}$.
\def\theequation{2.3.\arabic{subsubsection}.\arabic{equation}}
\subsubsection{\label{Non-BPS-Z=0-Spectra}Non-BPS, $Z=0$}

For the class $\mathcal{O}_{non-BPS,Z=0}$ of \textit{``large''} non-BPS, $%
Z=0 $ orbits the situation changes, and the result of \cite{TT} no longer
holds true, due to the local vanishing of $Z$.

In all \textit{magic} $N=2$, $d=4$ MESGT's the complex representation $%
R_{H_{0}}$ of $H_{0}$ decomposes under $\widetilde{h}^{\prime }=\frac{%
mcs\left( \widetilde{H}\right) }{U(1)}$ in the following way (see Eq. (\ref
{KR-KR-KR})):
\begin{equation}
R_{H_{0}}\longrightarrow \mathcal{W}_{\widetilde{h}^{\prime }}+\mathcal{Y}_{%
\widetilde{h}^{\prime }}+\mathbf{1}_{\mathbb{C}}\mathbf{,}  \label{deccomp}
\end{equation}
where in the r.h.s. the complex $\widetilde{h}^{\prime }$-singlet and the
complex non-singlet representations $\mathcal{W}_{\widetilde{h}^{\prime }}$
and $\mathcal{Y}_{\widetilde{h}^{\prime }}$ of $\widetilde{h}^{\prime }$
appear. Correspondingly, the decomposition of the $H_{0}$-invariant
representation $\left( R_{H_{0}}\right) ^{3}$ in terms of representations of
$\widetilde{h}^{\prime }$ reads (see Eq. (\ref{decomp-non-BPS-Z=0}))
\begin{equation}
\left( R_{H_{0}}\right) ^{3}\longrightarrow \left( \mathcal{W}_{\widetilde{h}%
^{\prime }}\right) ^{2}\mathcal{Y}_{\widetilde{h}^{\prime }}+\left( \mathcal{%
Y}_{\widetilde{h}^{\prime }}\right) ^{2}\mathbf{1}_{\mathbb{C}}\mathbf{.}
\label{decomp-decomp}
\end{equation}
Let us now recall that $dim_{\mathbb{R}}R_{H_{0}}=2n_{V}$ and $dim_{\mathbb{R%
}}\mathbf{1}_{\mathbb{C}}=2$, and let us define
\begin{equation}
\left.
\begin{array}{r}
dim_{\mathbb{R}}\mathcal{W}_{\widetilde{h}^{\prime }}\equiv \mathbf{W}_{%
\widetilde{h}^{\prime }}; \\
\\
dim_{\mathbb{R}}\mathcal{Y}_{\widetilde{h}^{\prime }}\equiv \mathbf{Y}_{%
\widetilde{h}^{\prime }};
\end{array}
\right\} :\mathbf{W}_{\widetilde{h}^{\prime }}+\mathbf{Y}_{\widetilde{h}%
^{\prime }}+2=2n_{V}.  \label{one}
\end{equation}

Thus, it can generally be stated that the mass spectrum along $\mathcal{O}%
_{non-BPS,Z=0}$ of all \textit{magic} $N=2$, $d=4$ symmetric MESGT's splits
under $\widetilde{h}^{\prime }=\frac{mcs(\widetilde{H})}{U(1)}$ as follows:

$\mathbf{-}$\textbf{\ }the mass ``modes'' fitting the $\mathbf{W}_{%
\widetilde{h}^{\prime }}$ real degrees of freedom corresponding to the
complex ($U(1)$-charged) non-$\widetilde{h}^{\prime }$-singlet
representation $\mathcal{W}_{\widetilde{h}^{\prime }}$ (which does \textit{%
not }couple to the complex $\widetilde{h}^{\prime }$-singlet in the $H_{0}$%
-invariant decomposition (\ref{decomp-decomp})) remain \textit{massless};

$\mathbf{-}$\textbf{\ }the mass ``modes'' fitting the $\mathbf{Y}_{%
\widetilde{h}^{\prime }}+2$ real degrees of freedom corresponding to the
complex ($U(1)$-charged) non-$\widetilde{h}^{\prime }$-singlet
representation $\mathcal{Y}_{\widetilde{h}^{\prime }}$ and to the ($U(1)$%
-charged) $\widetilde{h}^{\prime }$-singlet $\mathbf{1}_{\mathbb{C}}$
\textit{all} become \textit{massive}.

The resulting interpretation of the decomposition (\ref{deccomp}) is
\begin{equation}
R_{H_{0}}\longrightarrow \left(
\begin{array}{c}
\mathcal{W}_{\widetilde{h}^{\prime }} \\
\\
\mathbf{W}_{\widetilde{h}^{\prime }}~~\text{\textit{massless}}
\end{array}
\right) +\left(
\begin{array}{c}
\mathcal{Y}_{\widetilde{h}^{\prime }}+\mathbf{1}_{\mathbb{C}} \\
\\
\mathbf{Y}_{\widetilde{h}^{\prime }}+2~~\text{\textit{massive}}
\end{array}
\right) \mathbf{.}  \label{two}
\end{equation}

The interpretations (\ref{deccomp}) and (\ref{two}) show that, even though
the complex representations $\mathcal{W}_{\widetilde{h}^{\prime }}$, $%
\mathcal{Y}_{\widetilde{h}^{\prime }}$ and $\mathbf{1}_{\mathbb{C}}$ of $%
\widetilde{h}^{\prime }$ are charged with respect to the explicit factor $%
U(1)$ \textit{always} appearing in $\widetilde{h}$, this fact does \textit{%
not} affect in any way the splitting of the non-BPS, $Z=0$ mass ``modes''.

The critical mass spectra of the irreducible sequence $\frac{SU(1,1+n)}{%
U(1)\times SU(1+n)}$ and of the reducible sequence $\frac{SU(1,1)}{U(1)}%
\times \frac{SO(2,2+n)}{SO(2)\times SO(2+n)}$ are treated in Appendices I
and II of \cite{BFGM1}, respectively.\bigskip

Generally, the Hessian $\mathbf{H}^{V_{BH}}$ at \textit{regular} $N=2$,
non-BPS critical points of $V_{BH}$ exhibits the following features: it does
not have \textit{``repeller''} directions (i.e. strictly \textit{negative}
real eigenvalues), it has a certain number of \textit{``attractor''}
directions (related to strictly \textit{positive} real eigenvalues), but it
is also characterized by some \textit{vanishing} eigenvalues, corresponding
to massless non-BPS ``modes''.

\textit{A priori}, in order to establish whether the considered $N=2$,
non-BPS citical points of $V_{BH}$ are actually \textit{attractors} in a
strict sense, \textit{i.e.} whether they actually are \textit{stable minima}
of $V_{BH}$ in the scalar manifold, one should proceed further with
covariant differentiation of $V_{BH}$, dealing (at least) with third and
higher-order derivatives.

The detailed analysis of the issue of stability of both classes of \textit{%
regular} non-BPS critical points ($Z\neq 0$ and $Z=0$) of $V_{BH}$
in $N=2$, $d=4$ (symmetric) MESGT's was performed in
\cite{ferrara4}. In that paper it was found that, for \textit{all}
supergravities with homogeneous (not necessarily symmetric) scalar
manifolds the massless Hessian modes are actually \textit{``flat''}
directions of $V_{BH}$, \textit{i.e.} that the Hessian massless
modes persist, at the critical points of $V_{BH}$ itself, at all
order in covariant differentiation of $V_{BH}$. This is reported in
the next Section. \setcounter{equation}0
\def\theequation{2.\arabic{subsection}.\arabic{equation}}
\subsection{\label{Moduli-Spaces}From Massless Hessian Modes of $V_{BH}$%
\newline
to Moduli Spaces of Attractors}

In $N=2$ homogeneous (not necessarily symmetric) and $N>2$-extended (all
symmetric), $d=4$ supergravities the Hessian matrix of $V_{BH}$ at its
critical points is in general \textit{semi-positive definite}, eventually
with some vanishing eigenvalues (\textit{massless Hessian modes}), which
actually are \textit{flat} directions of $V_{BH}$ itself \cite
{Ferrara-Marrani-1,ferrara4}. Thus, it can be stated that for all
supergravities based on homogeneous scalar manifolds the critical points of $%
V_{BH}$ which correspond to \textit{``large''} black holes
(\textit{i.e.} for which one finds that  $V_{BH}\neq 0$) all are
\textit{stable}, up to some eventual \textit{flat} directions.

As pointed out above, the Attractor Equations of $N=2$, $d=4$ MESGT with $%
n_{V}$ Abelian vector multiplets may have \textit{flat} directions in the
non-BPS cases \cite{Ferrara-Marrani-1,ferrara4}, but \textit{not} in the $%
\frac{1}{2}$-BPS one \cite{FGK} (see Eqs. (\ref{BPS-crit}) and (\ref
{BPS-crit-2}) above).
\begin{table}[t]
\begin{center}
\begin{tabular}{|c||c|c|c|}
\hline
& $
\begin{array}{c}
\\
\frac{\widehat{H}}{\widehat{h}} \\
~
\end{array}
$ & $
\begin{array}{c}
\\
\text{r} \\
~
\end{array}
$ & $
\begin{array}{c}
\\
\text{\textit{dim}}_{\mathbb{R}} \\
~
\end{array}
$ \\ \hline\hline
$
\begin{array}{c}
\\
II:\mathbb{R}\oplus \Gamma _{n+2} \\
(n=n_{V}-3\in \mathbb{N\cup }\left\{ 0,-1\right\} )
\end{array}
$ & $SO(1,1)\times \frac{SO(1,n+1)}{SO(n+1)}~$ & $
\begin{array}{c}
\\
1(n=-1) \\
2(n\geqslant 0) \\
~
\end{array}
$ & $n+2~$ \\ \hline
$
\begin{array}{c}
\\
III:J_{3}^{\mathbb{O}} \\
~
\end{array}
$ & $\frac{E_{6(-26)}}{F_{4(-52)}}~$ & $2~$ & $6~$ \\ \hline
$
\begin{array}{c}
\\
IV:J_{3}^{\mathbb{H}} \\
~
\end{array}
$ & $\frac{SU^{\ast }(6)}{USp(6)}~$ & $2~$ & $14~$ \\ \hline
$
\begin{array}{c}
\\
V:J_{3}^{\mathbb{C}} \\
~
\end{array}
$ & $\frac{SL(3,C)}{SU(3)}~$ & $2~$ & $8~$ \\ \hline
$
\begin{array}{c}
\\
VI:J_{3}^{\mathbb{R}} \\
~
\end{array}
$ & $\frac{SL(3,\mathbb{R})}{SO(3)}~$ & $2~$ & $5$ \\ \hline
\end{tabular}
\end{center}
\caption{\textbf{Moduli spaces of non-BPS }$Z\neq 0$ \textbf{\ critical
points of } $V_{BH,N=2}$ \textbf{in }$N=2,d=4$ \textbf{symmetric
supergravities (}$\widehat{h}$\textbf{\ is the maximal compact subgroup of }$%
\widehat{H}$).\textbf{\ They are the} $N=2,d=5$ \textbf{symmetric real
special manifolds} \protect\cite{ferrara4} }
\end{table}

\begin{table}[h]
\begin{center}
\begin{tabular}{|c||c|c|c|}
\hline
& $
\begin{array}{c}
\\
\frac{\widetilde{H}}{\widetilde{h}}=\frac{\widetilde{H}}{\widetilde{h}%
^{\prime }\times U(1)} \\
~
\end{array}
$ & $
\begin{array}{c}
\\
\text{r} \\
~
\end{array}
$ & $
\begin{array}{c}
\\
\text{\textit{dim}}_{\mathbb{C}} \\
~
\end{array}
$ \\ \hline\hline
$
\begin{array}{c}
\\
I:Quadratic~~Sequence \\
(n=n_{V}-1\in \mathbb{N\cup }\left\{ 0\right\} ) \\
~
\end{array}
$ & $\frac{SU(1,n)}{U(1)\times SU(n)}~$ & $1~$ & $n~$ \\ \hline
$
\begin{array}{c}
\\
II:\mathbb{R}\oplus \Gamma _{n+2} \\
(n=n_{V}-3\in \mathbb{N\cup }\left\{ 0,-1\right\} )
\end{array}
$ & $\frac{SO(2,n)}{SO(2)\times SO(n)},n\geqslant 1~$ & $
\begin{array}{c}
\\
1(n=1) \\
2(n\geqslant 2) \\
~
\end{array}
$ & $n~$ \\ \hline
$
\begin{array}{c}
\\
III:J_{3}^{\mathbb{O}} \\
~
\end{array}
$ & $\frac{E_{6(-14)}}{SO(10)\times U(1)}~$ & $2~$ & $16~$ \\ \hline
$
\begin{array}{c}
\\
IV:J_{3}^{\mathbb{H}} \\
~
\end{array}
$ & $\frac{SU(4,2)}{SU(4)\times SU(2)\times U(1)}~$ & $2~$ & $8~$ \\ \hline
$
\begin{array}{c}
\\
V:J_{3}^{\mathbb{C}} \\
~
\end{array}
$ & $\frac{SU(2,1)}{SU(2)\times U(1)}\times \frac{SU(1,2)}{SU(2)\times U(1)}%
~ $ & $2~$ & $4~$ \\ \hline
$
\begin{array}{c}
\\
VI:J_{3}^{\mathbb{R}} \\
~
\end{array}
$ & $\frac{SU(2,1)}{SU(2)\times U(1)}~$ & $1~$ & $2~$ \\ \hline
\end{tabular}
\end{center}
\caption{\textbf{Moduli spaces of non-BPS }$Z=0$ \textbf{\ critical points
of } $V_{BH,N=2}$ \textbf{in }$N=2,d=4$ \textbf{symmetric supergravities (}$%
\widetilde{h}$\textbf{\ is the maximal compact subgroup of }$\widetilde{H}$)%
\textbf{. They are (non-special) symmetric K\"{a}hler manifolds}
\protect\cite{ferrara4}}
\end{table}

Tables 5 and 6 respectively list the moduli spaces of non-BPS $Z\neq 0$ and
non-BPS $Z=0$ attractors for symmetric $N=2$, $d=4$ special K\"{a}hler
geometries, for which a complete classification is available \cite{ferrara4}
(the attractor moduli spaces should exist also in homogeneous non-symmetric $%
N=2$, $d=4$\ special K\"{a}hler geometries, but their classification is
currently unknown). The general ``\textit{rule of thumb'' }to construct the
moduli space of a given attractor solution in the considered symmetric
framework is to coset the \textit{stabilizer} of the corresponding charge
orbit by its \textit{mcs}. By such a rule, the $\frac{1}{2}$-BPS attractors
do \textit{not} have an associated moduli space simply because the
stabilizer of their supporting BH charge orbit is \textit{compact}. On the
other hand, \textit{all} attractors supported by BH charge orbits whose
stabilizer is \textit{non-compact} exhibit a non-vanishing moduli space.
furthermore, it should be noticed that the non-BPS $Z\neq 0$ moduli spaces
are nothing but the symmetric real special scalar manifolds of the
corresponding $N=2$, $d=5$ supergravity.\smallskip

Nevertheless, it is worth remarking that some symmetric $N=2$, $d=4 $
supergravities have no non-BPS \textit{flat }directions at all.

The unique $n_{V}=1$ symmetric models are the so-called $t^{2}$ and $t^{3}$
models; they are based on the rank-$1$ scalar manifold $\frac{SU\left(
1,1\right) }{U\left( 1\right) }$, but with different holomorphic
prepotential functions.

The $t^{2}$ model is the first element ($n=0$) of the sequence of
irreducible symmetric special K\"{a}hler manifolds $\frac{SU\left(
1,n+1\right) }{U\left( 1\right) \times SU\left( n+1\right) }$ ($n_{V}=n+1$, $%
n\in \mathbb{N\cup }\left\{ 0\right\} $) (see \textit{e.g.} \cite{BFGM1} and
Refs. therein), endowed with \textit{quadratic} prepotential. Its bosonic
sector is given by the $\left( U\left( 1\right) \right) ^{6}\rightarrow
\left( U\left( 1\right) \right) ^{2}$ truncation of
Maxwell-Einstein-axion-dilaton (super)gravity, \textit{i.e.} of \textit{pure}
$N=4$, $d=4$ supergravity (see \textit{e.g.} \cite{FHM} and \cite{CFGM-1}
for recent treatments).

On the other hand, the $t^{3}$ model has \textit{cubic} prepotential; as
pointed out above, it is an \textit{isolated case} in the classification of
symmetric SK manifolds (see \textit{e.g.} \cite{CFG}; see also \cite
{LA08-Proc} and Refs. therein), but it can be thought also as the $\mathit{%
s=t=u}$\textit{\ degeneration }of the $stu$ model. It is worth pointing out
that the $t^{2}$ and $t^{3}$ models are based on the same rank-$1$ SK
manifold, with different constant \textit{scalar curvature}, which
respectively can be computed to be (see \textit{e.g.} \cite{BFM-SIGRAV06}
and Refs. therein)
\begin{equation}
\begin{array}{l}
\frac{SU(1,1)}{U(1)},~t^{2}~\text{\textit{model}}:R=-2; \\
\\
\frac{SU(1,1)}{U(1)},\text{~}t^{3}~\text{\textit{model}}:R=-\frac{2}{3}.
\end{array}
\end{equation}
Beside the $\frac{1}{2}$-BPS attractors, the $t^{2}$ model admits only
non-BPS $Z=0$ critical points of $V_{BH}$ with no \textit{flat} directions.
Analogously, the $t^{3}$ model admits only non-BPS $Z\neq 0$ critical points
of $V_{BH}$ with no \textit{flat} directions.

For $n_{V}>1$, the non-BPS $Z\neq 0$ critical points of $V_{BH}$, if any,
all have \textit{flat} directions, and thus a related moduli space (see
Table 5). However, models with no non-BPS $Z=0$ \textit{flat} directions at
all and $n_{V}>1$ exist, namely they are the first and second element ($n=-1$%
, $0$) of the sequence of reducible symmetric special K\"{a}hler manifolds $%
\frac{SU\left( 1,1\right) }{U\left( 1\right) }\times \frac{SO\left(
2,n+2\right) }{SO\left( 2\right) \times SO\left( n+2\right) }$ ($n_{V}=n+3$,
$n\in \mathbb{N\cup }\left\{ 0,-1\right\} $) (see \textit{e.g.} \cite{BFGM1}
and Refs. therein), \textit{i.e.} the so-called $st^{2}$ and $stu$ models,
respectively. The $stu$ model (\cite{DLR,BKRSW}, see also \textit{e.g.} \cite
{stu-unveiled}\ and Refs. therein) has two non-BPS $Z\neq 0$ \textit{flat}
directions, spanning the moduli space $SO\left( 1,1\right) \times SO\left(
1,1\right) $ (\textit{i.e.} the scalar manifold of the $stu$ model in $d=5$%
), but \textit{no} non-BPS $Z=0$ \textit{massless Hessian modes} at all. On
the other hand, the $st^{2}$ model (which can be thought as the $\mathit{t=u}
$\textit{\ degeneration} of the $stu$ model) has one non-BPS $Z\neq 0$
\textit{flat} direction, spanning the moduli space $SO\left( 1,1\right) $ (%
\textit{i.e.} the scalar manifold of the $st^{2}$ model in $d=5$), but
\textit{no} non-BPS $Z=0$ \textit{flat} direction at all. The $st^{2}$ is
the \textit{``smallest''} symmetric model exhibiting a non-BPS $Z\neq 0$
\textit{flat} direction.

Concerning the \textit{``smallest''} symmetric models exhibiting a non-BPS $%
Z=0$ \textit{flat} direction they are the second ($n=1$) element of the
sequence $\frac{SU\left( 1,n+1\right) }{U\left( 1\right) \times SU\left(
n+1\right) }$ and the third ($n=1$) element of the sequence $\frac{SU\left(
1,1\right) }{U\left( 1\right) }\times \frac{SO\left( 2,n+2\right) }{SO\left(
2\right) \times SO\left( n+2\right) }$. In both cases, the unique non-BPS $%
Z=0$ \textit{flat} direction spans the non-BPS $Z=0$ moduli space $\frac{%
SU\left( 1,1\right) }{U\left( 1\right) }\sim \frac{SO\left( 2,1\right) }{%
SO\left( 2\right) }$ (see Table 6), whose local geometrical properties
however differ in the two cases (for the same reasons holding for the $t^{2}$
and $t^{3}$ models treated above).

\begin{table}[t]
\begin{center}
\begin{tabular}{|c||c|c|c|}
\hline
& $
\begin{array}{c}
\\
\frac{1}{N}\text{-BPS orbits } \frac{G}{\mathcal{H}} \\
~
\end{array}
$ & $
\begin{array}{c}
\\
\text{non-BPS, }Z_{AB}\neq 0\text{ orbits}~\frac{G}{\widehat{\mathcal{H}}}
\\
~
\end{array}
$ & $
\begin{array}{c}
\\
\text{non-BPS, }Z_{AB}=0\text{ orbits }\frac{G}{\widetilde{\mathcal{H}}} \\
\\
~
\end{array}
$ \\ \hline\hline
$
\begin{array}{c}
\\
N=3 \\
~
\end{array}
$ & $\frac{SU(3,n)}{SU(2,n)}~$ & $-$ & $\frac{SU(3,n)}{SU(3,n-1)}~$ \\ \hline
$
\begin{array}{c}
\\
N=4 \\
~
\end{array}
$ & $SU(1,1)\times \frac{SO(6,n)}{SO(2)\times SO(4,n)}~$ & $SU(1,1)\times
\frac{SO(6,n)}{SO(1,1)\times SO(5,n-1)}~$ & $SU(1,1)\times \frac{SO(6,n)}{%
SO(2)\times SO(6,n-2)}$ \\ \hline
$
\begin{array}{c}
\\
N=5 \\
~
\end{array}
$ & $\frac{SU(1,5)}{SU(3)\times SU\left( 2,1\right) }$ & $-$ & $-$ \\ \hline
$
\begin{array}{c}
\\
N=6 \\
~
\end{array}
$ & $\frac{SO^{\ast }(12)}{SU(4,2)}~$ & $\frac{SO^{\ast }(12)}{SU^{\ast }(6)}%
~$ & $\frac{SO^{\ast }(12)}{SU(6)}~$ \\ \hline
$
\begin{array}{c}
\\
N=8 \\
~
\end{array}
$ & $\frac{E_{7\left( 7\right) }}{E_{6\left( 2\right) }}$ & $\frac{%
E_{7\left( 7\right) }}{E_{6\left( 6\right) }}$ & $-~$ \\ \hline
\end{tabular}
\end{center}
\caption{\textbf{\textit{``Large''} charge orbits of the real, symplectic }$%
R_{V}$ \textbf{representation of the }$U$\textbf{-duality group }$G$ \textbf{%
supporting BH attractors with non-vanishing entropy in $N\geqslant 3$%
-extended, $d=4$ supergravities} \textbf{(}$n$\textbf{\ is the number of
matter multiplets) }\protect\cite{review-Kallosh}}
\end{table}

\begin{table}[t]
\begin{center}
\begin{tabular}{|c||c|c|c|}
\hline
& $
\begin{array}{c}
\\
\frac{1}{N}\text{-BPS} \\
\text{moduli space }\frac{\mathcal{H}}{\frak{h}}\text{ } \\
~
\end{array}
$ & $
\begin{array}{c}
\\
\text{non-BPS, }Z_{AB}\neq 0 \\
\text{moduli space }\frac{\widehat{\mathcal{H}}}{\widehat{\frak{h}}} \\
~
\end{array}
$ & $
\begin{array}{c}
\\
\text{non-BPS, }Z_{AB}=0 \\
\text{moduli space }\frac{\widetilde{\mathcal{H}}}{\widetilde{\frak{h}}} \\
~
\end{array}
$ \\ \hline\hline
$
\begin{array}{c}
\\
N=3 \\
~
\end{array}
$ & $\frac{SU(2,n)}{SU(2)\times SU\left( n\right) \times U\left( 1\right) }~$
& $-$ & $\frac{SU(3,n-1)}{SU(3)\times SU\left( n-1\right) \times U\left(
1\right) }~$ \\ \hline
$
\begin{array}{c}
\\
N=4 \\
~
\end{array}
$ & $\frac{SO(4,n)}{SO(4)\times SO\left( n\right) }~$ & $SO(1,1)\times \frac{%
SO(5,n-1)}{SO(5)\times SO\left( n-1\right) }~$ & $\frac{SO(6,n-2)}{%
SO(6)\times SO\left( n-2\right) }$ \\ \hline
$
\begin{array}{c}
\\
N=5 \\
~
\end{array}
$ & $\frac{SU\left( 2,1\right) }{SU\left( 2\right) \times U\left( 1\right) }
$ & $-$ & $-$ \\ \hline
$
\begin{array}{c}
\\
N=6 \\
~
\end{array}
$ & $\frac{SU(4,2)}{SU(4)\times SU\left( 2\right) \times U\left( 1\right) }~$
& $\frac{SU^{\ast }(6)}{USp\left( 6\right) }~$ & $-$ \\ \hline
$
\begin{array}{c}
\\
N=8 \\
~
\end{array}
$ & $\frac{E_{6\left( 2\right) }}{SU\left( 6\right) \times SU\left( 2\right)
}$ & $\frac{E_{6\left( 6\right) }}{USp\left( 8\right) }$ & $-~$ \\ \hline
\end{tabular}
\end{center}
\caption{\textbf{Moduli spaces of BH attractors with non-vanishing entropy
in $N\geqslant 3$-extended, $d=4$ supergravities (}$\frak{h}$\textbf{, }$%
\widehat{\frak{h}}$\textbf{\ and }$\widetilde{\frak{h}}$\textbf{\ are
maximal compact subgroups of }$\mathcal{H}$\textbf{, }$\widehat{\mathcal{H}}$%
\textbf{\ and }$\widetilde{\mathcal{H}}$\textbf{, respectively, and }$n$
\textbf{is the number of matter multiplets)} \protect\cite{review-Kallosh}}
\end{table}

We conclude by recalling that in \cite{GZ,GZMcR-1,GZMcR-2} it was shown that
the $N=2$, $d=5$ \textit{magic} MESGT's defined by $J_{3}^{\mathbb{C}%
},J_{3}^{\mathbb{H}}$ and $J_{3}^{\mathbb{O}}$ are simply the
``lowest'' members of three infinite families of unified $N=2$,
$d=5$ MESGT's defined by Lorentzian Jordan algebras of degree $>3$.
The scalar manifolds of such theories are not homogeneous except for
the ``lowest'' members. It would be interesting to extend the
analysis of \cite{FG2} and \cite{BFGM1} to these theories in five
dimensions and to their descendants in $d=5$, respectively.
\setcounter{equation}0
\def\theequation{\arabic{section}.\arabic{equation}}
\section{\label{N>2,d=4}$U$-Duality \textit{``Large''} Orbits\newline
and Moduli Spaces of Attractors\newline
in $N\geqslant 3$-Extended, $d=4$ Supergravities}

In $N\geqslant 3$-extended, $d=4$ supergravities, whose scalar manifold is
always symmetric, there are \textit{flat} directions of $V_{BH}$ at both its
BPS and non-BPS critical points. As mentioned above, from a
group-theoretical point of view this is due to the fact that the
corresponding supporting BH charge orbits always have a \textit{non-compact}
stabilizer \cite{ferrara4,review-Kallosh}. The BPS \textit{flat} directions
can be interpreted in terms of left-over hypermultiplets' scalar degrees of
freedom in the truncation down to the $N=2$, $d=4$ theories \cite
{ADF-U-duality-d=4,Ferrara-Marrani-1}. In Tables 7 and 8 all (classes of)
\textit{``large''} charge orbits and the corresponding moduli spaces of
attractor solution in $N\geqslant 3$-extended, $d=4$ supergravities are
reported \cite{review-Kallosh}.

\section{\label{Conclusions}Conclusions}

In the present report we dealt with results holding at the
classical, Einstein supergravity level. It is conceivable that the
\textit{flat} directions of classical extremal BH attractors will
be removed (\textit{i.e.} lifted) by \textit{quantum}
(\textit{perturbative} and \textit{non-perturbative}) corrections
(such as those coming from higher-order derivative contributions
to the gravity and/or gauge sector) to the \textit{classical}
effective BH potential $V_{BH}$. Consequently, \textit{at the
quantum} \textit{level, moduli spaces for attractor solutions may
not exist at all} (and therefore also \textit{the actual
attractive nature of the critical points of }$V_{BH}$\textit{\
might be destroyed}). \textit{However, this may not be the case
for }$N=8$.

In the presence of \textit{quantum} lifts of \textit{classically
flat} directions of the Hessian matrix of $V_{BH}$ at its critical
points, in order to answer the key question: \textit{``Do extremal
BH attractors (in a strict sense) survive at the quantum
level?''}, it is thus crucial to determine whether such lifts
originate from Hessian modes with \textit{positive} squared mass
(corresponding to \textit{attractive} directions) or with
\textit{negative} squared mass (\textit{i.e.} \textit{tachyonic}, \textit{%
repeller} directions).

The fate of the unique non-BPS $Z\neq 0$ flat direction of the $st^{2}$
model in presence of the most general class of quantum perturbative
corrections consistent with the axionic-shift symmetry has been studied in
\cite{Quantum-Lift}, showing that, as intuitively expected, the \textit{%
classical solutions get lifted at the quantum level}.
Interestingly, in \cite {Quantum-Lift}\textbf{\ }it is found that
the \textit{quantum} lift occurs more often towards
\textit{repeller} directions (thus destabilizing the whole
critical solution, and \textit{destroying the attractor in a
strict sense}), than towards \textit{attractive}
directions.\textbf{\ }The same behavior may be expected for the
unique non-BPS $Z=0$\ flat direction of the $n=2$\ element of the
quadratic irreducible sequence and the $n=3$\ element of the cubic
reducible sequence (see above).

Generalizing it to the presence of more than one \textit{flat}
direction, this would mean that \textit{only a (very) few
classical attractors do remain attractors in a strict sense at the
quantum level}; consequently, \textit{at
the quantum} (\textit{perturbative} and \textit{non-perturbative}) \textit{%
level the ``landscape'' of extremal BH attractors should be strongly
constrained and reduced}.\medskip

Despite the considerable number of papers written on the
\textit{Attractor Mechanism} in the extremal BHs of the
supersymmetric theories of gravitation in past years, still much
remains to be discovered along the way leading to a deep
understanding of the inner dynamics of (eventually extended)
space-time singularities in supergravities, and hopefully of their
fundamental high-energy counterparts, such as $d=10$ superstrings
and $d=11$ $M$-theory.

\section*{Acknowledgments}

This work is supported in part by the ERC Advanced Grant no. 226455, \textit{%
``Supersymmetry, Quantum Gravity and Gauge Fields''}
(\textit{SUPERFIELDS}).

A. M. would like to thank the \textit{William I. Fine Theoretical Physics
Institute} (FTPI) of the University of Minnesota, Minneapolis, MN USA, the
\textit{Center for Theoretical Physics} (CTP) of the University of
California, Berkeley, CA USA, and the Department of Physics and Astronomy of
the University of California, Los Angeles, CA USA, where part of this work
was done, for kind hospitality and stimulating environment. Furthermore, A.
M. would like to thank Ms. Hanna Hacham for peaceful and inspiring
hospitality in Palo Alto, CA USA.

The work of S. B. ~has been supported in part by the grant
INTAS-05-7928.

The work of S. F.~has been supported also in part by INFN - Frascati
National Laboratories, and by D.O.E.~grant DE-FG03-91ER40662, Task
C.

The work of M. G. ~has been supported in part by National Science
Foundation under grant number PHY-0555605. Any opinions, findings
and conclusions or recommendations expressed in this material are
those of the authors and do not necessarily reflect the views of the
National Science Foundation.

The work of A. M. has been supported by an INFN visiting Theoretical
Fellowship at SITP, Stanford University, Stanford, CA, USA.

\end{document}